\documentclass[aps,prb,twocolumn,groupedaddress]{revtex4-1}
\usepackage{graphicx}
\usepackage{float}
\usepackage{amsmath,amsfonts,amssymb,mathtools}
\usepackage{multirow}

\def\code#1{\texttt{#1}}

\begin{document}



\title{Beyond theory driven discovery: hot random search and datum derived structures}

\author{Chris J.\ Pickard} \email[]{cjp20@cam.ac.uk}
\affiliation{Department of Materials Science \& Metallurgy, University
  of Cambridge, 27 Charles Babbage Road, Cambridge CB3~0FS, United
  Kingdom} \affiliation{Advanced Institute for Materials Research,
  Tohoku University 2-1-1 Katahira, Aoba, Sendai, 980-8577, Japan}


\date{\today}

\begin{abstract}
   
Data driven methods have transformed the prospects of the computational chemical sciences, with machine learned interatomic potentials (MLIPs) speeding up calculations by several orders of magnitude. I reflect on theory driven, as opposed to data driven, discovery based on \emph{ab initio} random structure searching (AIRSS), and then introduce two methods which exploit machine learning acceleration. I show how long high throughput anneals, between direct structural relaxation, enabled by ephemeral data derived potentials (EDDPs), can be incorporated into AIRSS to bias the sampling of challenging systems towards low energy configurations. Hot AIRSS (hot-AIRSS) preserves the  parallel advantage of random search, while allowing much more complex systems to be tackled. This is demonstrated through searches for complex boron structures in large unit cells. I then show how low energy carbon structures can be directly generated from a single, experimentally determined, diamond structure. An extension to the generation of random \emph{sensible}  structures, candidates are stochastically generated and then optimised to minimise the difference between the EDDP environment vector and that of the reference diamond structure. The distance-based cost function is captured in an actively learned EDDP.  Graphite, small nanotubes and caged, fullerene-like, structures emerge from searches using this potential, along with a rich variety of tetrahedral framework structures. Using the same approach, the pyrope, Mg$_3$Al$_2$(SiO$_4$)$_3$, garnet structure is recovered from a low energy AIRSS structure generated in a smaller unit cell with a different chemical composition. The relationship of this approach to modern diffusion model based generative methods is discussed.
\end{abstract}


\maketitle

\section{Introduction}

The introduction of unbiased, first principles, structure prediction in the mid-2000s revolutionised materials discovery.\cite{oganov2019structure} It was no longer necessary to trawl through databases of the “usual suspects”, or to concoct novel structures by hand. Unknown structure types, and surprising phenomena, emerged from explorations of the density functional theory (DFT) energy landscape, where previous approaches to structure prediction depended on the fast evaluation of empirical forcefields.\cite{stillinger1985computer,biswas1985interatomic,tersoff1988empirical,woodley1999prediction} DFT provides an approximation to the underlying quantum mechanical interactions governing the stability of different phases, balancing computational efficiency with a robustness\cite{lejaeghere2016reproducibility} that permits genuine predictions. In Section \ref{theory} I will highlight several examples of theory driven discovery.

There is a new revolution underway, sparked by the discovery that machine learning techniques can routinely be exploited to accelerate the exploration of energy landscapes, either through molecular dynamics (MD) or structure prediction. From early attempts in the 1990s,\cite{brown1996combining} the groundbreaking contributions of Behler\cite{behler2007generalized} and Csanyi\cite{bartok2010gaussian} have stimulated the development of a wide array of machine learned interatomic potentials (MLIPs).\cite{bartok2017machine}Among these are the ephemeral data derived potentials (EDDPs)\cite{pickard2022ephemeral,salzbrenner2023developments} - see Section \ref{EDDP} - which were introduced with the explicit aim of accelerating \emph{ab initio} random structure search (AIRSS).\cite{pickard2006high,pickard2011ab} In Section \ref{HotAIRSS} I will show how the multiple order of magnitude acceleration offered by EDDPs over DFT allow for a style of calculation that would have simply been too computationally expensive previously - hot-AIRSS, the integration of long MD driven anneals as part of the high throughput optimisation of stochastically generated structures. Finally, in Section \ref{measure} and \ref{vectors} I show how to extend the concept of generating random sensible structures - see Section \ref{sensible}, to the point of being very closely related to modern diffusion model based generative approaches.

\section{Theory driven discovery}
\label{theory}

AIRSS \cite{pickard2006high,pickard2011ab} is built on the high throughput first principles relaxation of diverse stochastically generated structures (from crystals, to clusters, molecules, surfaces, interfaces, and grain boundaries). The emphasis is on exploration, and the hunting for outliers, or surprises, through an attempt to uniformly sample configuration space, within a defined distribution of candidate structures. 

Throughout my work there is a focus on the discovery of unexpected phenomena, as opposed to the detail of a particular crystal structure - not forgetting that it is essential that the structural details are correctly identified in order meaningfully predict the discovered material's properties. When a surprising result is encountered, considerable effort is expended in attempting to identify the competing phases that might render the prediction unsound. In many cases this is indeed the outcome. Persisting in this approach leads to a high success rate, with few false positives, and high-quality predictions.

The first applications of AIRSS were to the to the high-pressure sciences, beginning with an exploration of superconductivity and metallicity in the dense hydrides.\cite{pickard2006high,pickard2007metallization} This has grown to be a very active area with many well-known successes\cite{pickard2020superconducting} - see Section \ref{htpsc}. With other first principles structure prediction techniques\cite{oganov2019structure} - USPEX,\cite{oganov2006crystal} CALYPSO,\cite{wang2012calypso} and XtalOpt,\cite{lonie2011xtalopt} AIRSS is now a key tool for materials discovery with applications ranging from battery materials\cite{lu2021ab,zhu2021accelerating} to molecular polymorphism,\cite{smalley2022structure} and nanoconfined water.\cite{kapil2022first} 

The emphasis of first principles random structure search on highly parallelisable and broad sampling ensures it is particularly well adapted to modern computational trends, statistical physics and machine learning in particular, where it has become an indispensable source of training data.\cite{deringer2018data,merchant2023scaling,zeni2023mattergen}

\subsection{Mixed phases in hydrogen}

An early application of AIRSS was an attempt to understand Phase III of dense hydrogen, and in particular identifying model structures that exhibited the key vibrational spectroscopic signatures measured in diamond anvil cell experiments.\cite{pickard2007structure} Our prediction of the C2/c-24 structure as the best model for phase III is standing the test of time.\cite{loubeyre2020synchrotron,monacelli2023quantum} 

Analysing the large number of AIRSS generated structures I was confronted by a striking family of metastable structures, of a type not that had not previously suggested for an element. They consisted of layers, alternating between graphene-like and molecular, see Figure \ref{structures}. I felt these structures must be important and potentially dynamically stabilised phases (either through zero-point motion, or temperature), but the techniques were not then ready to allow a full phase diagram to be computed. Nevertheless, we published the mixed phase structures in Ref. \onlinecite{pickard2007structure} and emphasised them in presentations to experimentalists. 

Initially the mixed phases did not address any open experimental questions and were largely ignored. This changed when Goncharov and Gregoryanz approached me with a puzzle - they were seeing a surprising softening in a high frequency Raman peak in warm (room temperature) hydrogen at megabar pressures. I suggested that they were observing a mixed phase, and on investigation this  proved to be the case.\cite{howie2012mixed} The mixed phases are now an established feature of the hydrogen phase diagram. It is fair to say that, given the experimental challenges in determining the positions of protons, our current understanding of dense hydrogen is largely due to first principles structure searches, with much having been mapped out in Ref. \onlinecite{pickard2007structure}. 

Why was first principles structure search so successful in tackling this well explored problem? Of course, the high throughput nature of the searches made a big difference, increasing the sheer number of structures considered. But the most important structures could probably have been found using contemporary MD methods. The fact that they were not is likely because MD was frequently conducted in cubic, or orthorhombic, unit cells, and with fixed numbers of atoms, typically multiples of 8. But my candidates for dense hydrogen, C2/c-24, Cmca-12 and the mixed phases, all contained multiples of 12 atoms. I had been in the habit of not assuming the number of atoms in the unit cell and choosing them randomly as part of the structure generation. This was also to be very important for  aluminium, described below in Section \ref{altpa}, and highlights the importance of minimally biased stochastic searches.

\subsection{Ionic ammonia}

\begin{figure}
\label{theorydiscovery}
\includegraphics[width=0.45\textwidth]{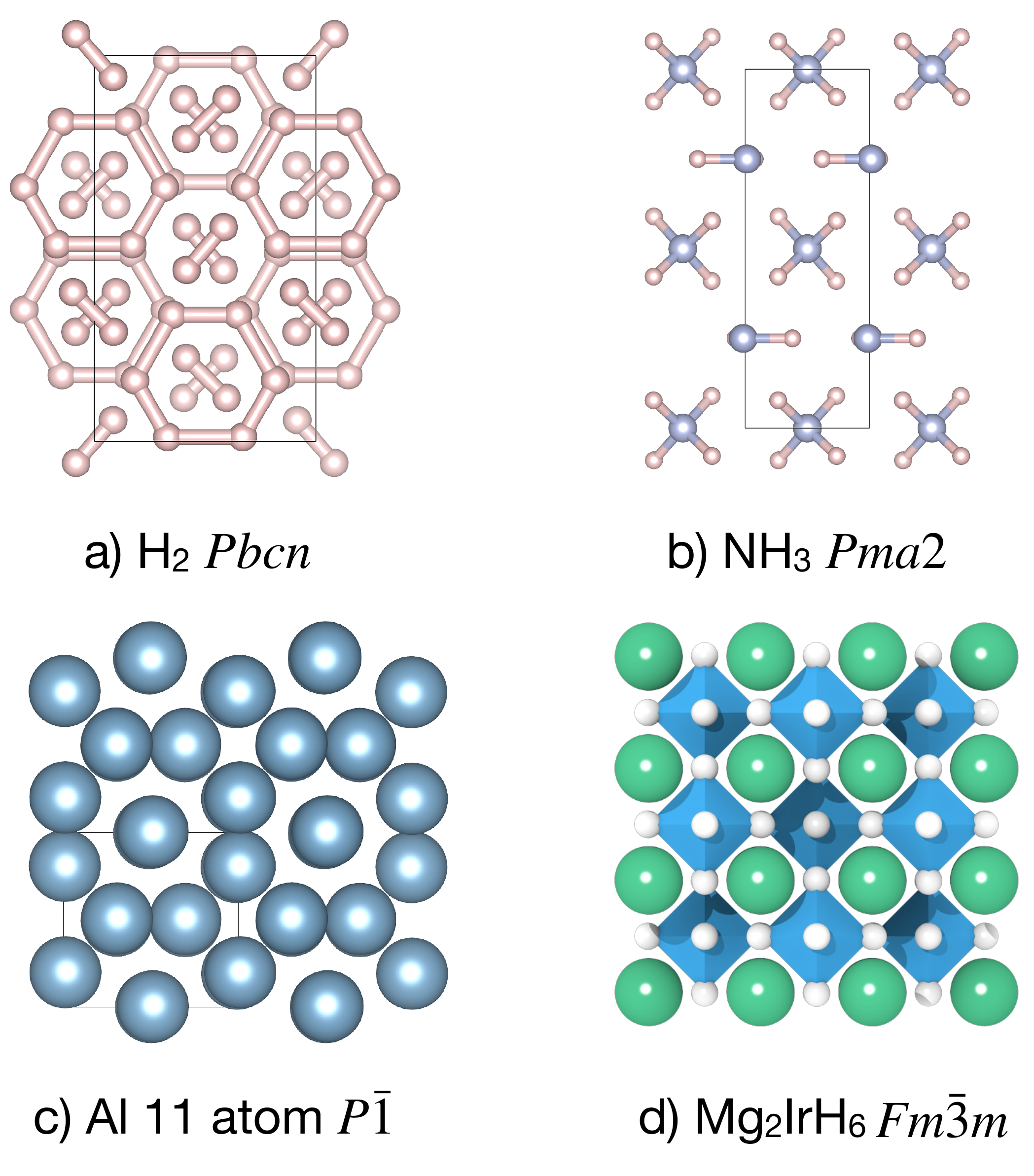}
\caption{a) Pbcn mixed phase of hydrogen at 300~GPa, b) Pma2 NH$_2$-NH$_4$ phase of ammonia at 100~GPa, c) 11-atom host guest phase of aluminium at 5~TPa, and d) dynamically stable 0~GPa cubic phase of Mg$_2$IrH$_6$ with predicted superconducting T$_c$ of 160~K.}
\label{structures}
\end{figure}

When searching for molecular crystal structures, a well-established protocol is to stochastically pack connected molecular units.\cite{pickard2011ab,price2014predicting} This shrinks the search space, as compared to a less restricted search starting from unconnected atoms, and dramatically increases the odds of finding low energy configurations. But it is at the cost of potentially missing the most stable one, if it does not adhere to the chosen molecular unit. In the spirit of assuming as little as it is computationally feasible to, I had been searching for dense phases of NH$_3$ by randomly placing the N and H atoms into randomly shaped unit cells individually. It was a routine project, but I was jolted awake one early morning while checking the results of the overnight runs. The most stable units under pressure were, by some margin, NH$^-_2$ and NH$^+_4$ - see Figure \ref{structures}, not the expected NH$_3$.\cite{pickard2008highly} I assumed that something was wrong with the calculations. This possibility had not been discussed for pure ammonia previously, and it was not something we were looking for. After careful testing, the result held, and the spontaneous ionisation of NH$_3$ has been experimentally established.\cite{ninet2014experimental} Spontaneous self-ionisation more generally is now considered as a possibility where it might not have been previously. 

\subsection{Complex phases of aluminium at terapascal pressures}
\label{altpa}

We (and others, particularly Yanming Ma and co-workers) had starting to find a great number of electride type structures in the dense elements.\cite{pickard2009dense,ma2009transparent} One striking feature of these were the localisation of states under increasing pressure, and band narrowing. I wondered whether I could find a non-magnetic element that under the right conditions would exhibit magnetism. I began the hunt, systematically working my way through the periodic table. Importantly, it turned out, I was randomly choosing the number of atoms in the unit cell. When it came to aluminium, I was surprised to find the most stable structure at 3 TPa contained 11 atoms in the unit cell. At that time few groups would even consider odd numbers of atoms as a possibility, based on the heuristic that they would unlikely be the most stable. The 11-atom cell was, however, significantly more stable than the other candidates, and initially, when I visualised it, it made no sense. It appeared to be amorphous, or still random somehow. This was unusual, as the most stable structures usually exhibit some symmetry. But I continued building supercells and spinning the structure around in the visualiser, and eventually all became clear. 

The structure consisted of tubes and chains of atoms - see Figure \ref{structures}. I was aware of the work of Nelmes and McMahon\cite{mcmahon2006high} on incommensurate host guest phases in the alkali metals\cite{mcmahon2004incommensurate} as Volker Heine had publicised it in the Theory of Condensed Matter Group, Cambridge. This turned out to be exactly what I was seeing in the 11-atom structure - an approximant of a kind of 1D quasicrystal. Once I had recognised that, it was straightforward to manually construct other, larger, approximants, and estimate the ideal lattice parameters for the host and guest phases. I was also able to determine that the structure was of the electride type and construct a simple model for it,\cite{pickard2010aluminium} based on a generalised Lennard-Jones model, which later became the basis of the EDDPs - see Section \ref{EDDP}.

This result has not been confirmed experimentally - yet. But it has had an impact on the field - it showed that materials under extreme compression might be complex, and not just simply close packed. This has inspired the high-pressure community, particularly the shock physicists, for example being used as part of the justification for using the National Ignition Facility (NIF) to perform exploratory science.\cite{gorman2022experimental} Continuing my sweep through the periodic table, I did eventually manage to find magnetism in an electride phase, in potassium.\cite{pickard2011predicted}

\subsection{High throughput hunt for conventional superconductivity}
\label{htpsc}

Bringing the applications of AIRSS up-to-date, recent work has refocussed on the search for high temperature superconductors, specifically the hydrides, which may be (meta-)stable at ambient pressures, and superconduct at temperatures exceeding the critical temperature (T$_c$) of magnesium diboride. The field of hydride superconductivity has not been without controversy,\cite{garisto2024superconductivity} and it is essential to be able to identify candidate superconductors that might maybe be synthesised at low pressures, opening the field to broad and intense experimental scrutiny.

With the growth of computational resources since the debut of AIRSS, as well as refinements in the methods and optimisations of the key DFT code used for structural optimisation (CASTEP\cite{clark2005first}), it is now possible to add an additional layer of sampling to the searches. While early studies would concentrate on elements or compounds with a fixed composition, it later became possible to study the composition space of a given binary, or ternary, system.\cite{conway2021rules,nelson2021navigating} The next step has been to search over a wide range of composition spaces simultaneously, in a high throughput manner. 

In an initial study we  explored the binary hydrides over a range of pressures from 100 GPa to 500 GPa.\cite{shipley2021high} Several novel superconducting hydrides were discovered, and known ones rediscovered. The maximum superconducting transition temperatures, T$_c$, varied from 380 K at 500 GPa, to above 250 K at 100 GPa. A striking feature of our result was that the T$_c$ did not drop precipitously as the pressure was reduced, and through extrapolation one might expect hydride T$_c$s to be as high as 200 K at ambient pressures. This stimulated an extension of this approach to the ternary hydrides at low, and ambient pressure.\cite{dolui2024feasible} 

The searches across composition space were performed entirely using first principles methods - and so theory driven at this stage, and resulted in the discovery of Mg$_2$IrH$_6$ as a dynamically stable, moderately metastable, candidate conventional superconductor with a predicted T$_c$ of 160~K. Once Mg$_2$IrH$_6$ had been identified, detailed structure searches over the Mg-Ir-H composition space, accelerated with the EDDP machine learned interatomic potentials (see Section \ref{EDDP}) provided a thorough picture of the competing phases, as well as a feasible synthesis route. Having highlighted the power of theory driven search for discovery, this most recent work touches its limits, and demonstrates the power of data driven approaches, which will be the focus of the rest of this contribution.

\section{Generating random sensible structures}
\label{sensible}


Key to the success of AIRSS is the initial step of generating an ensemble of chemically \emph{sensible} random structures for subsequent high throughput structure relaxation. This step is performed by the \code{buildcell} code of the GPL2 open source AIRSS package.\cite{AIRSS-website} The random structures are constructed once an appropriate distribution of parameters has been selected - based on either chemical insights or previous calculations (see Section \ref{measure}). When building a random unit cell its volume and shape should be chosen. These must be selected from a range, and it makes sense to choose this range to adhere experimentally reasonable values - even if only very approximately so. There is little point in searching in excessively small, or large, unit cells. Similar choices must be made for other parameters - how closely should atoms be permitted to approach each other in the initial structures? Structures might be generated to have randomly generated space (or point) group symmetries. The structural units might be molecules or fragments, rather than individual atoms. Composition can be stochastically chosen, but the ranges of compositions to be considered must be specified. Some thought should be given to \emph{load balancing} the searches - each of the stochastically generated structures should have roughly the same computational cost.

The initial random structures look \emph{sensible} and certainly some of them might be expected to have reasonably low energies, even before structural optimisation. Put together, these choices define a \emph{generative} model, in machine learning terminology. This will be explored further in Sections \ref{measure} and \ref{vectors}, and the relation to modern generative approaches to structure prediction will be discussed in Section \ref{discuss}.

\section{Ephemeral data derived potentials}
\label{EDDP}

The prospect for data derived potentials to accelerate structure search had long been apparent, and in Ref. \onlinecite{deringer2018data} it was shown the random structure search and gaussian approximation potentials (GAP)\cite{bartok2010gaussian} could be combined to iteratively generate a robust boron potential. At that time, the development of GAP potentials was  relatively intricate and time consuming, and the resulting potentials slow. To ensure the AIRSS could routinely benefit from the promised acceleration, with minimal interruption to the successful high throughput workflow, ephemeral data derived potentials (EDDPs) were introduced.\cite{pickard2022ephemeral} The emphasis on their \emph{ephemeral} nature was intended to draw the attention away of the difficult task of developing high-quality benchmarked potentials, towards the generation of disposable potentials that could be trained and used rapidly.

EDDPs are based on a simple model for the interatomic interaction, inspired by Lennard-Jones style potentials, with a minimal extension to handle many body interactions.\cite{pickard2022ephemeral,salzbrenner2023developments} The resulting feature, or environment, vectors are the input for small neural networks (in many cases, a single hidden layer with just five nodes). Multiple neural networks are fit, in parallel with random initialisations, just as in AIRSS. Early stopping, based on a validation portion of the 80:10:10 training:validation:testing data split,\cite{prechelt1998early} is used to discourage overfitting. The Levenberg-Marquadt (LM) optimiser is found to be fast and produce excellent training and testing losses. Combining the many neural networks together, minimising the non-negative least squared (NNLS) error, again to the validation split, results in a sparse ensemble, with only a fraction of the neural networks being selected for the final model. The ensemble enables the variance of the predicted energies among the many fits to be evaluated, and this can be used to detect pathological structures, as well as to drive an active learning to less certain configurations.\cite{hansen1990neural,schran2020committee}

A key feature of EDDPs is that they are trained on the DFT energies of large numbers of small, and so rapid to compute, structures. To date, forces are not used in the training, which might be a limitation compared to other methods. However, there are advantages to this approach, and using AIRSS to generate many highly diverse structures the resulting potentials have proven to be more than adequate for the purposes of accelerating structure prediction. In Ref. \onlinecite{salzbrenner2023developments} it is shown that EDDPs can also be used as the basis for reliable and quantitative molecular and lattice dynamics simulations. The structures encountered in a random search are extremely varied as compared to those sampled by molecular dynamics, and this diversity of the structures on which the EDDPs are trained appears to largely eliminate the problems of stability of molecular dynamics simulations. 

EDDPs have been extended to be able to handle large numbers of chemical species using the alchemical ideas of Cerriotti.\cite{lopanitsyna2023modeling} The GPL2 open source EDDP package is available.\cite{EDDP-website}

\section{Hot Random Structure Search}
\label{HotAIRSS}

For many problems AIRSS is an extremely effective approach to discovering low energy structures. The first principles potential energy surface is relatively smooth, and for moderate system sizes the probability of encountering low energy configurations is sufficiently high that when coupled with high throughput computation AIRSS is a competitive structure prediction technique.\cite{pickard2011ab} However, as more complex problems are attempted, the exponential growth in local minima begins to dominate, and without extensive use of constraints to prepare sensible initial starting points the likelihood of generating low energy configurations becomes too low to justify the computational effort in searching for them. For example, in Ref. \onlinecite{pickard2022ephemeral}, an EDDP was generated for boron, and a free search for $\gamma$-boron\cite{oganov2009ionic} was attempted. No symmetry was exploited, nor was the knowledge that boron tends to favour icosahedra, and unit cells containing 28 boron atoms at approximately the correct density were generated. A slightly distorted version of the orthorhombic \emph{Pnnm} $\gamma$-boron structure was successfully located, but only twice out of 362 754 putative structures. In tests, the 12 atom $\alpha$-boron structure can typically be found in free AIRSS searches once every 3000 attempts. Making an assumption of an exponential increase in difficulty, we might estimate that identifying the $\gamma$-boron structure in a doubled cell of 56 atoms would take something like $3\times10^6$ structure optimisations, unfeasible from first principles and challenging even using EDDPs. 

A difficulty that the use of fast potentials for structure search has created is the management and storage of the vast number of structures that can be generated on even modest computer hardware. The writing of the data to disk can become a bottleneck on some high-performance computing (HPC) systems. One option is to only store the most stable structures encountered, for example by rejecting any new structures that are outside a given energy threshold of any previously encountered for that composition. An alternative is to embrace the acceleration and perform more intense computation for each generated and stored structure.

\begin{figure}
\includegraphics[width=0.3\textwidth]{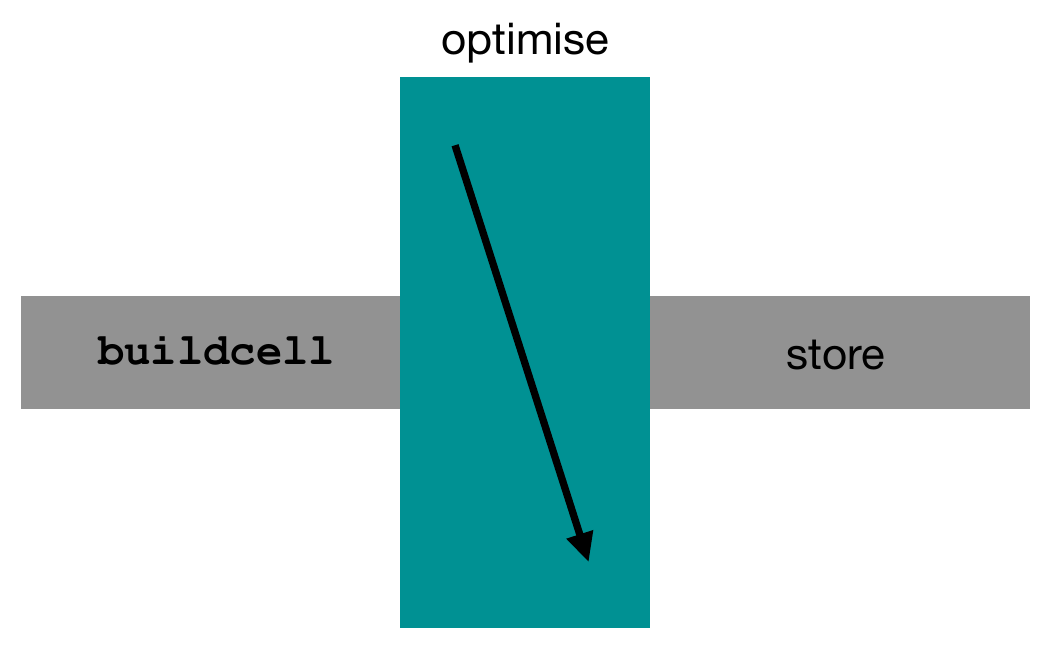}
\includegraphics[width=0.45\textwidth]{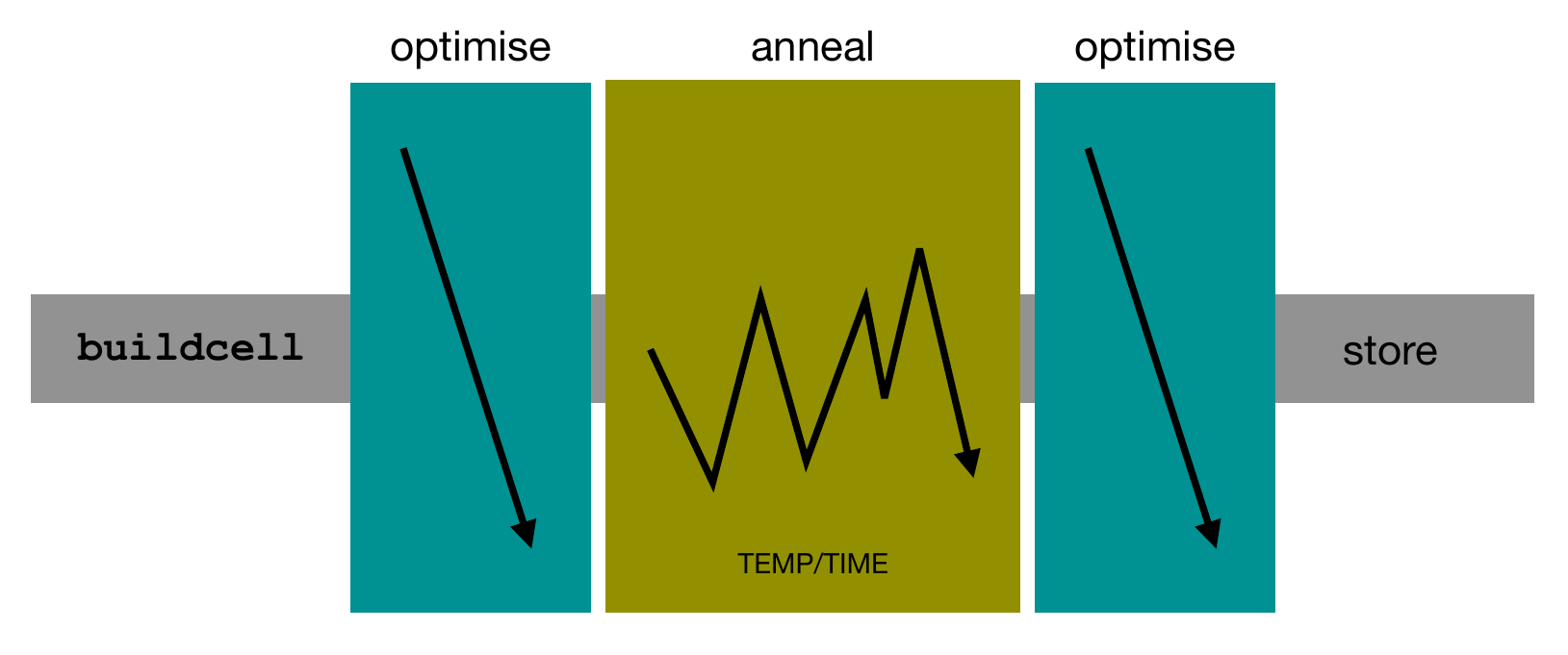}
\caption{\emph{Top:} A representation of AIRSS:  a random sensible structure is generated using the \code{buildcell} code, and is then structurally optimised to the nearest local minimum of the energy landscape, which is either described by DFT, or a fast equivalent, such as an EDDP. The resulting structure is stored. This is repeated, in parallel, a large number of times. \emph{Bottom:} hot-AIRSS proceeds in a similar manner, but after the first optimisation with an EDDP, a long anneal is performed at a chosen temperature, close to but below the melting temperature, for a given time. The resulting structure is finally structurally optimised and stored.}
\label{AIRSSvHotAIRSS}
\end{figure}

Probably the greatest impact of the MLIP revolution has been the opening up of the possibility of performing long time-scale, large length-scale, MD simulations at approaching first principles quality.\cite{cheng2020evidence,schran2021machine,deringer2021origins,kapil2022first} We exploit this here to perform random structure search integrating an extended annealing period, between local optimisations. AIRSS, and what we term hot-AIRSS, are contrasted in Fig.\ref{AIRSSvHotAIRSS} An initial random structure is generated, just as in traditional AIRSS, potentially using the several strategies to prepare the structures described in Section \ref{sensible}, and relaxed to its nearest local minimum using the \code{repose} code. Rather than stopping there, the \code{ramble} molecular dynamics code supplied in the EDDP package is used to perform an anneal at a fixed temperature for a given time. The resulting structure is then again relaxed to the now nearest local minimum, which if the temperature chosen is sufficiently high is not likely to be the same as the initial one. 

The two parameters introduced are the temperature for the anneal (typically chosen to be approaching but below the melting temperature of the system), and the time for the anneal. The time is typically selected to exceed 10 picoseconds, and potentially as long as nanoseconds. There is no quenching of the system during the molecular dynamics run, and the overall process, given the final local optimisation, can be thought of as an elaborate optimisation scheme, and from the point of view of AIRSS is a direct replacement of the usual local optimiser. From this perspective it is reasonable to permit the exploitation of symmetry during the anneal. The \code{ramble} code implements symmetrised MD, a functionality that is not generally available in more widely used codes. While not currently implemented, the ability to optimise and run dynamics on defined structural units is likely to prove useful.

To explore the capability of hot-AIRSS we revisit the high-pressure phases of boron and attempt to locate the Pnnm $\gamma$-boron phase at 10~GPa. An EDDP is prepared so that the required high throughput MD driven anneals are feasible. It is generated using the \code{chain} script, with seven iterations of active learning. In the first step 10000 random structures containing 12, 24 and 28 boron atoms, are constructed, and their PBE GGA\cite{perdew1996generalized} single point energies are computed using CASTEP\cite{clark2005first} using the default QC5 OTFG pseudopotential for boron, a k-point spacing of 0.07 $2\pi$ \AA$^{-1}$, plane wave cutoff of 340 eV, and default grid scales. Marker structures consisting of 11 known and putative phases of boron are added to the dataset, each one shaken 1000 times with an amplitude of 0.1. For each iteration of active learning, AIRSS is used to generate 10000 structures at a randomly chosen pressure between 5~GPa and 15~GPa, which are each shaken once with an amplitude of 0.1. 30 individual potentials are trained, with NNLS selecting 12. The resulting training and testing MAE are 13.33 and 13.67 meV/atom respectively. 

\begin{figure}
\includegraphics[width=0.45\textwidth]{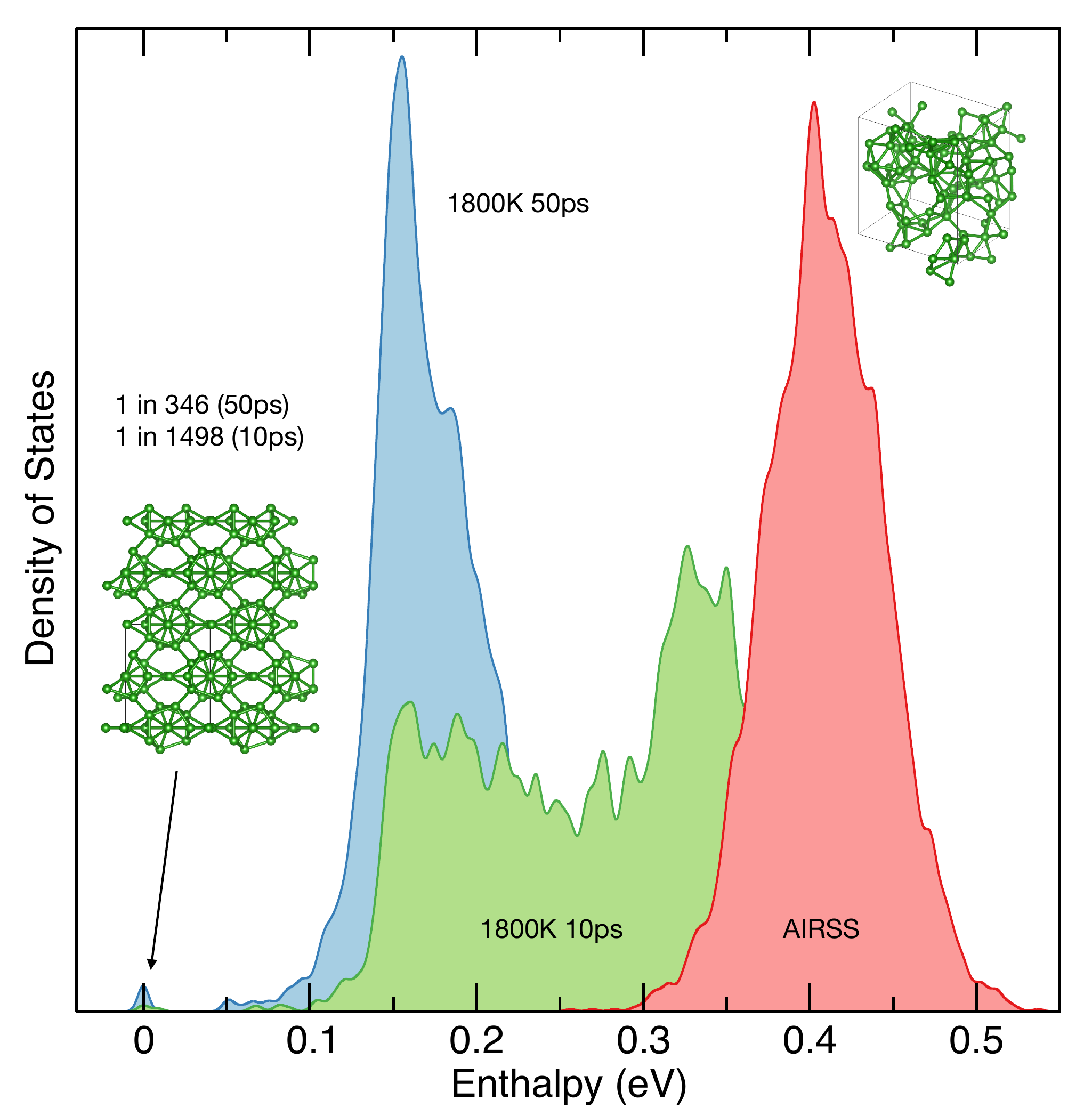}
\caption{\emph{Unconstrained search for 56 boron atoms at 10GPa} Structural densities of states for (red) an AIRSS search, (green) a hot-AIRSS search at 1800~K for 10~ps, and (blue) a hot-AIRSS search at 1800~K for 50~ps. The  enthalpy per boron atom relative to the ground-state Pnnm $\gamma$-boron phase (shown) is plotted.} 
\label{boron}
\end{figure}

The results of three searches for 56 atoms of boron at 10~GPa are presented in Figure \ref{boron}. The structures generated using traditional AIRSS are highly disordered. The most stable are around 0.3~eV/atom less stable than the known ground state $\gamma$-boron structure. The probability of generating low energy structures is low, and consistent with the above estimate of the difficulty of this task. Even given the very rapid structural optimisation this is not a viable approach to finding the ground state structure in such a large unit cell.

In the second search, hot-AIRSS is performed. After an initial relaxation, a 10~ps anneal at 1800~K is performed. This temperature is selected after conducting a few short runs and assessing the average mobility of atoms in the unit cell. The temperature should be below the melting temperature, as fully molten configurations relax to approximately the same distribution as AIRSS. However, the atoms should be sufficiently energetic so as to be mobile enough to explore a wide range of configurations. Should a low energy configuration be encountered, since the system is at below the melting temperature, it is liable to freezing. This is acceptable, since on further relaxation the low energy configuration will be maintained. In principle it should be possible to set the anneal temperature automatically, and on a per-sample basis, but this is not explored further here. 

The resulting structural density of states exhibits a much broader distribution, with an increased diversity of structures. Out of 2996 samples, two of the structures located are found to be identical to the known $\gamma$-boron structure. One of them was the 56 atom Pbcn modification of $\gamma$-boron discussed in Ref. \onlinecite{ahnert2017revealing}. On increasing the time of the anneal to 50~ps the distribution shifts to lower energies still, and the $\gamma$-boron phase is found 11 times out of 3806 samples. It should be noted that while the probability of encounter has increased by 4.3 each anneal was five times longer - so the length of anneal is a parameter that should be adjusted to maximise computational efficiency.

\begin{figure}
\includegraphics[width=0.45\textwidth]{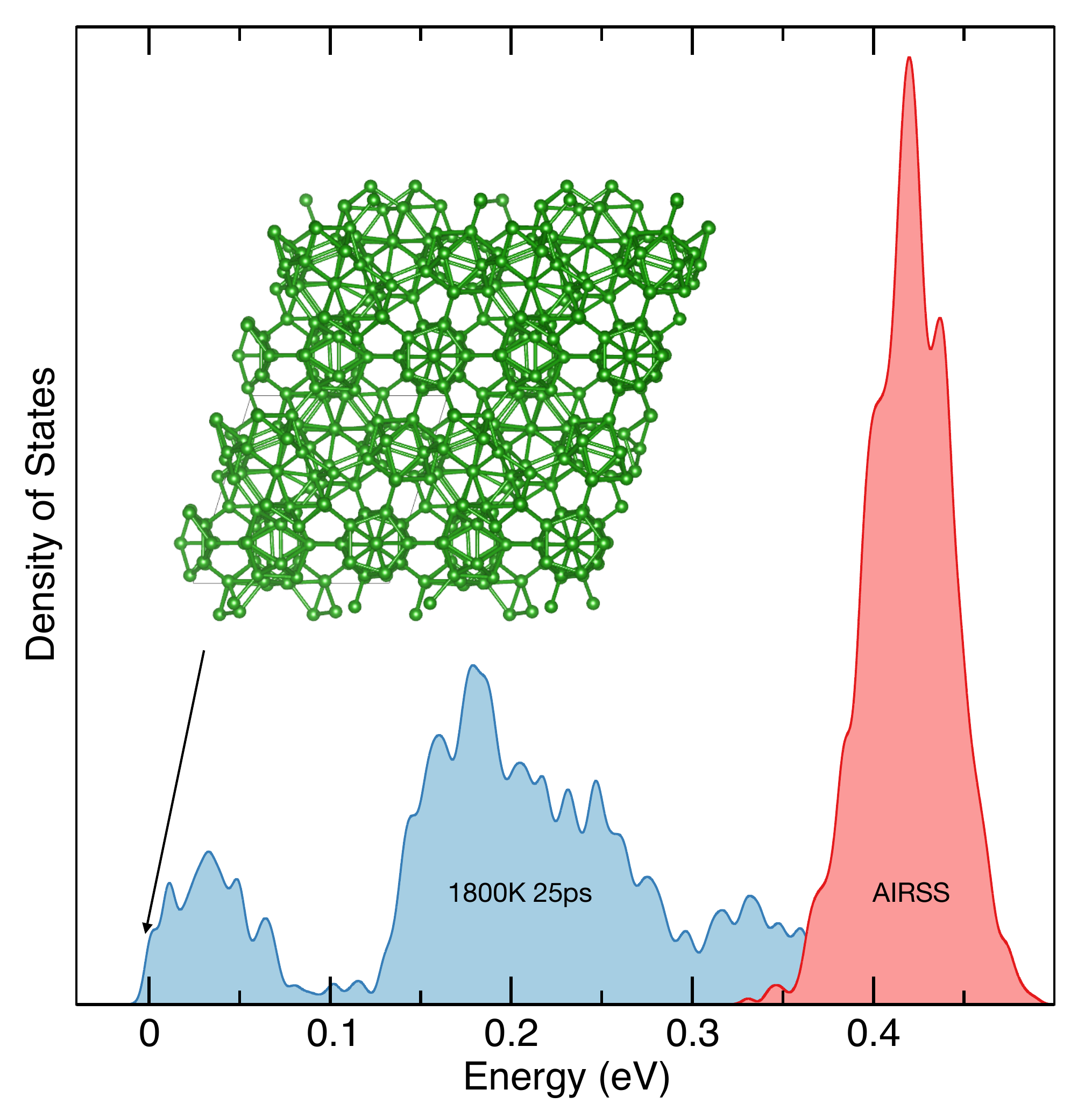}
\caption{\emph{Fixed cell search for 105 to 111 boron atoms} Structural densities of states for (red) an AIRSS search, (blue) a hot-AIRSS search at 1800~K for 25~ps. The energy per boron atom relative to the most stable structure (shown) is plotted. The lattice parameters for rhombohedral $\beta$-boron were fixed and taken from Ref. \onlinecite{callmer1977accurate}.}
\label{boron105}
\end{figure}

It is currently thought that rhombohedral $\beta$-boron is the most stable phase at low temperatures and pressures. The structure is complex, and likely highly defected leading to entropic stabilisation.\cite{van2007thermodynamic} In Ref. \onlinecite{deringer2018data} we used an actively learned GAP potential to explore the relative energy of the defects and interstitials. In Ref. \onlinecite{podryabinkin2019accelerating} it was shown that moment tensor potential\cite{shapeev2016moment} accelerated evolutionary algorithms could generate low energy approximants of rhombohedral $\beta$-boron without recourse to experimental information. Tetrahedral $\beta$-boron is thought to have a region of stability at elevated temperatures and pressures. Similarly to the rhombohedral phase, the tetrahedral phase is complex, with the best models containing 192 atoms in the primitive unit cell, and is also stabilised by a propensity to defect and interstitial formation. The stabilisation of these, and other, phases of boron have recently been studied in detail by Hayami \emph{et al.}\cite{hayami2024thermodynamic}.

\begin{figure}
\includegraphics[width=0.45\textwidth]{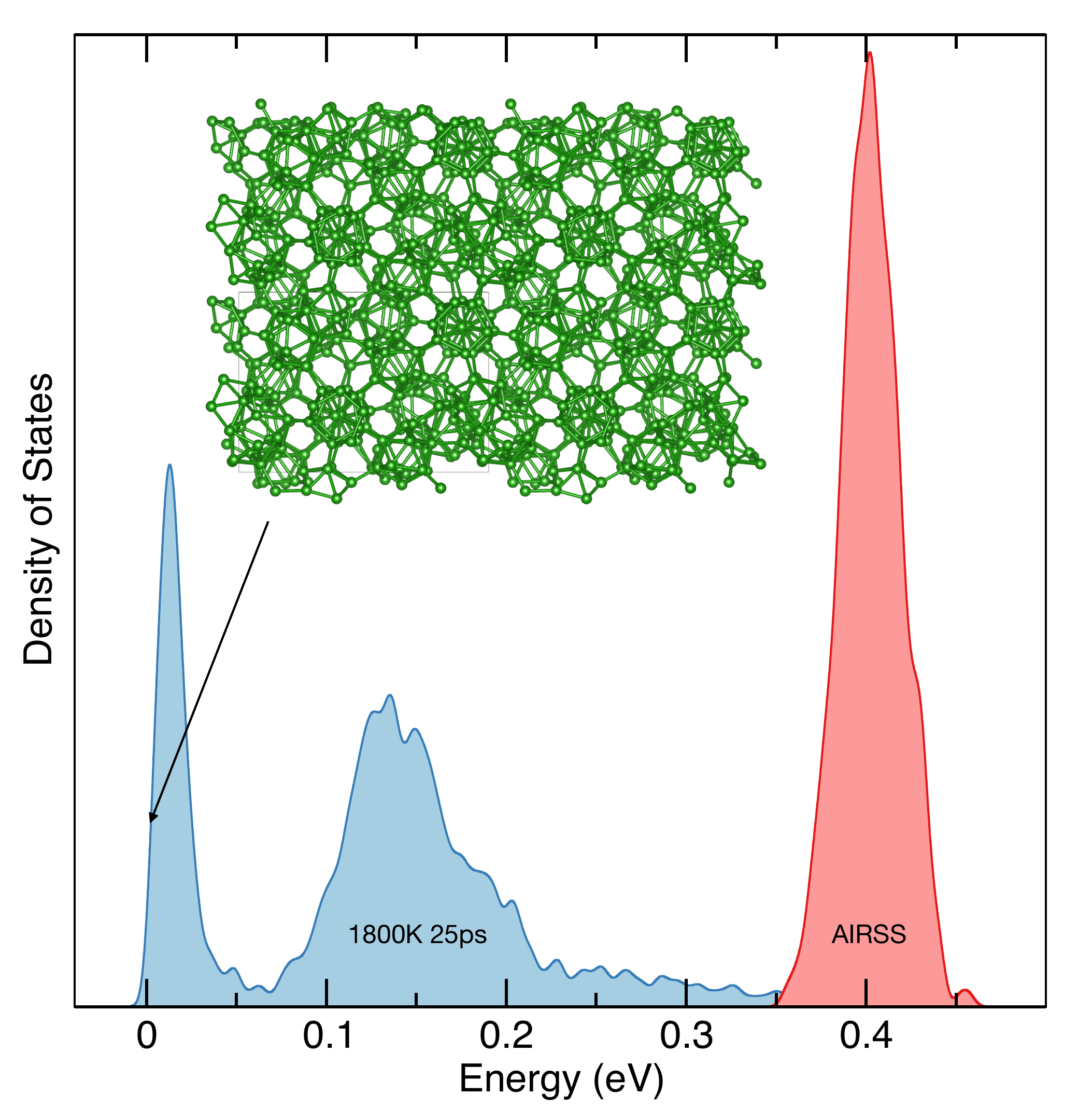}
\caption{\emph{Fixed cell search for 192 boron atoms} Structural densities of states for (red) an AIRSS search, (blue) a hot-AIRSS search at 1800~K for 25~ps. The energy per boron atom relative to the most stable structure (shown) is plotted. The lattice parameters for tetrahedral $\beta$-boron were fixed and taken from Ref. \onlinecite{hayami2015structural}.}
\label{boron192}
\end{figure}

In Figure \ref{boron105} the results of AIRSS and hot-AIRSS searches for 105 to 111 boron atoms in a single rhombohedral unit cell, fixed to experimental lattice parameters.\cite{callmer1977accurate} The density of structural states for the AIRSS search is narrowly peaked around 0.4~eV above the most stable structure found. The distribution of states from hot-AIRSS calculations at 1800~K for 25~ps is much broader, extending to lower energy. There is a peak at low energy, consisting of many structures visually similar to known $\beta$-boron models, but exhibiting a wide range of defects and interstitials, which can be expected to contribute to entropic stabilisation. The situation for tetragonal $\beta$-boron is very similar - see Figure \ref{boron192} - although the low energy peak of defective structures is significantly narrower in energy. Apart from the work of Podryabinkin \emph{et al.}, theoretical studies of the $\beta$-borons have proceeded by analysing defect and interstitial populations of the experimental structures. Here we see that hot-AIRSS can discover the underlying structural motifs of these complex phases.

hot-AIRSS is an elegant modification to AIRSS that maintains the trivial parallelisability of random structure search, and requires minimal changes to the computational workflow, or the provided \code{airss.pl} script in which the workflow is embodied. Temperature has been long recognised as a key parameter in structure search, most notably in simulated annealing,\cite{doll2008structure} basin hopping,\cite{wales1997global} and more explicitly through short molecular dynamics explorations in minima hopping.\cite{goedecker2004minima} The computationally efficient EDDPs now allow temperature to play a role in random structure search, and it is shown to be a powerful approach to tame complex and challenging systems.

\section{Generating structures from measured minimum separations}
\label{measure}
The computational creation of random, yet chemically sensible, structures is central to the success of AIRSS, see Section \ref{sensible}. One of the most powerful approaches is the building of structures satisfying a defined (but potentially stochastically generated) species-wise matrix of minimum separations - the MINSEP method of the AIRSS \code{buildcell} code. With the method additionally tagged with AUTO, the minimum separations are measured from the most stable structure with the desired composition, if available, along with a target density. If there are no structures available, the specified minimum separation parameters are used. 

For well packed inorganic materials the random structures generated in this way are likely to be chemically sensible and hence of relatively low energy when computed using DFT. The measured structures are typically the result of earlier, less constrained, searches. However, should experimentally known crystal structures be available for a given composition, the separations and density can be measured from those.

\section{Generating structures from measured feature vectors}
\label{vectors}

The development of many body descriptors, or feature/environment vectors, as the basis for MLIPs, such as the EDDPs described in Section \ref{EDDP}, open the way to much more sophisticated measurements to be made of atomistic structures. Related to the measurement of the minimum atomic separations, these descriptors provide a detailed measurement and description of the environment around a chosen atomic site. If structures can be generated that have similar environment vectors to a known, stable, structure then those structures are likely to be chemically similar to the target, and similarly low lying in the potential energy landscape. 

If the so generated structures exhibit some diversity, and are not identical to the target, this provides an alternative approach to building structures for AIRSS, and one might expect them to be not only sensible, but close to their nearby local minimum, and hence require little or no structural optimisation using DFT. Computing the single point total energies should be sufficient to rank the candidates.

\begin{figure}
\includegraphics[width=0.45\textwidth]{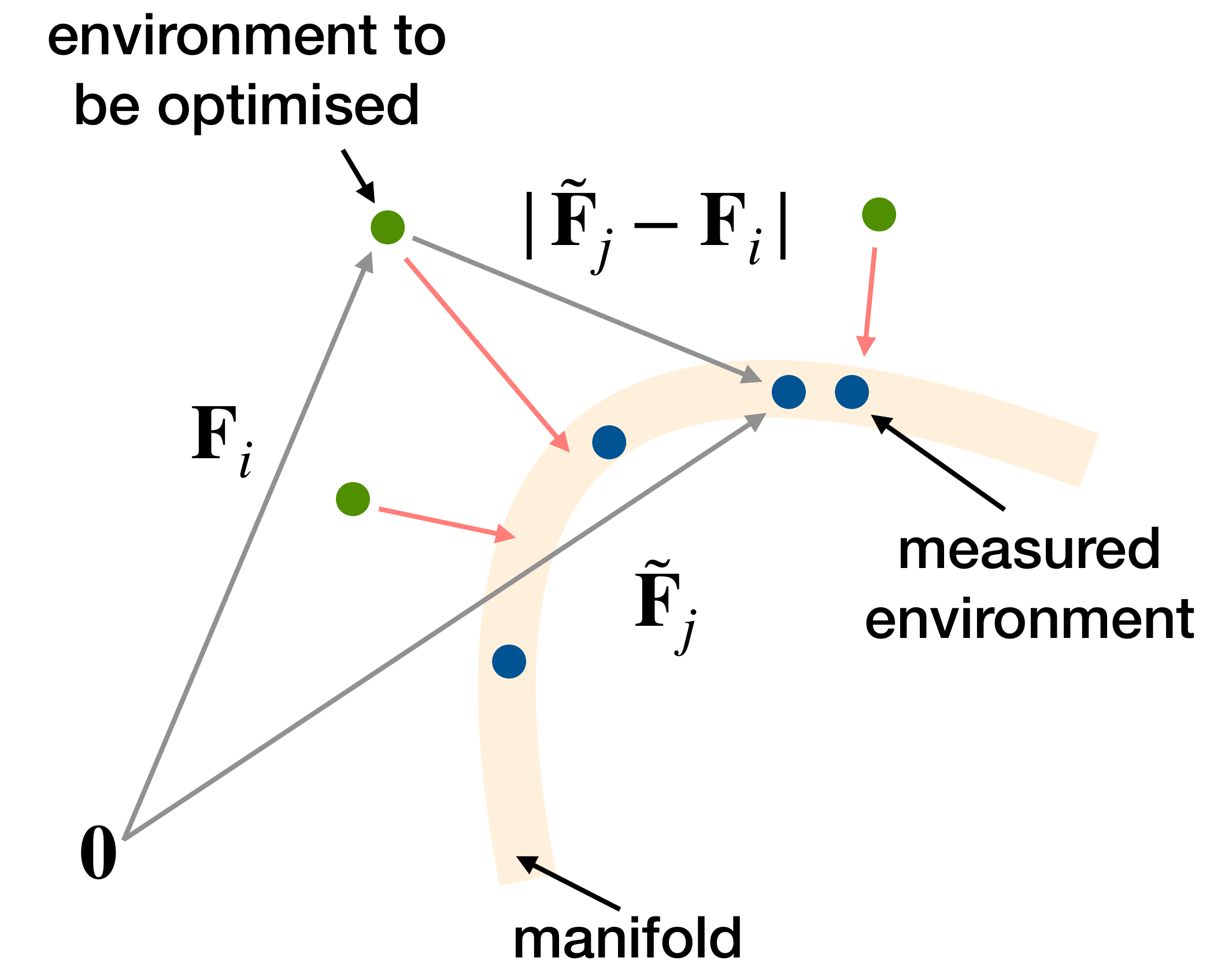}
\caption{\emph{Optimisation to the manifold of measured environments.} The blue circles are the environments, $j$, in the chosen feature space, measured from the target structure. They are assumed to lay on a low dimensional manifold embedded sketched by the light red band. The green circles represent the distinct environments, $i$, of the structure to be generated by optimisation towards the manifold, in the direction of the red arrows.}
\label{manifold}
\end{figure}

We now present such a scheme to generate structures that are closely related to a target structure. First, the feature vectors for the atomic environments in the target structure are computed. We will use the EDDP feature vectors, and these are obtained using the \code{frank} code. One might then perform an AIRSS search where the structural optimiser (for example, CASTEP in first principles searches and \code{repose} when EDDPs are used to accelerate the search) is replaced with a code that computes the gradient with respect to atomic displacements and changes in unit cell shape, of some cost function that monotonically depends on the distance of the new structure's feature vectors from the target vectors, see Figure \ref{manifold}. Here we instead actively train an EDDP on this cost function, using a modified version of the \code{chain} script, \code{manifest}. While a less direct approach, it has advantages.

Firstly, it permits the use of the AIRSS/EDDP tools with no modification - once the cost-based EDDP has been trained it can be used as any other EDDP, permitting structure searches using \code{repose}, molecular dynamics using \code{ramble} and lattice dynamics through \code{wobble}. Secondly, while the cost function may (or may not) be a strictly smooth function, the learned EDDP will be, by construction.

As the \code{manifest} script progresses, structures are generated either randomly, as in the first step of the iterative training of an EDDP, as shakes of the target structure (a marker structure), and from shaken AIRSS structures with intermediate generations of the cost-based potential. Instead of computing the DFT single point total energies for these configurations, the cost for each one is computed from the sum of a function of the distances from the configuration environments to the target environments. The training of the cost based EDDP then progresses iteratively, and rapidly as no DFT computations are required.

The cost contribution of a single environment in a structure is defined as a function of the soft minimum Euclidean distance to the potentially many environments of the target structure. This choice avoids the need to assign and pair the environments between the structure and the target structure and means that a minimum cost can be achieved if the environments of the new structure match any combination of the environments in the target structure.

A choice of the function of the Euclidean distance might be the commonly used squared distance. However, this function becomes very large for dissimilar environments, and the optimisation scheme may lose discrimination between environments similar to the target once the EDDP has been learned from the cost data. To maintain resolution close to the target environments the partial costs are evaluated as:

\begin{equation}
c_{ij}=\ln(\frac{\beta}{N^2}|{\bf F}_i-{\bf \tilde F}_j|^2+1).
\end{equation}

\noindent For small distances between the feature vectors ${\bf F}_i$ and ${\bf \tilde F}_j$, of length $N$, the squared Euclidean distance is recovered, but for large distances the cost is moderated, and does not grow to be too large. The parameter $\beta$ controls the degree to which small distances increase the cost, and so for large $\beta$ the cost is minimised by more strictly enforcing similarity with the target environments.

To evaluate the cost for each configuration, with respect to the target environments, the most straight forward approach is to identify the minimum partial cost for each atom in the configuration:

\begin{equation}
E_{\rm cost}=\sum_i \min_{j} \{c_{ij}\}.
\end{equation}

\noindent This approach has the disadvantage that the resulting cost landscape is not smooth. To some extent this could be managed through learning the EDDP representation of the cost landscape. However, it is preferable to instead construct a softened approximation to the minimum:

\begin{equation}
E_{\rm cost}=-\sum_i \frac{1}{\alpha}\ln\left( \frac{1}{M}\sum_j^M e^{-\alpha c_{ij}}\right),
\end{equation}

\noindent where $M$ is the total number of target environments. The parameter $\alpha$ controls the degree of softness of the approximation. For large values of $\alpha$ the strict minimum is recovered. It is worth noting that for typical values of $\alpha$ the cost for the target structures computed against themselves do not evaluate to zero. However, if $\alpha$ is appropriately set the cost should increase for all distortions of the target.

\section{Datum driven discovery}
\label{datum}

We have discussed the power of theory driven discovery in Section \ref{theory}. Data driven approaches are emerging as powerful methods to accelerate search and discovery, but it is instructive to consider what can be learned from a single data point, or datum. Using the scheme described above we first investigate the discovery potential of a using a single, experimentally known, structure as a generative source of hypothetical structures. We then explore how the approach might be integrated within a first principles searching strategy.

\subsection{Carbon}

Carbon is a fascinating element, with a great number of theoretically proposed allotropes,\cite{hoffmann2016homo} and fewer iconic experimentally known structures. Graphite is the thermodynamically favoured structure at ambient conditions, with diamond becoming stable at high pressures, and an important metastable material. At higher pressures still several phase transitions have been predicted, from \emph{bc8},\cite{yin1984si} to \emph{sc},\cite{grumbach1996phase} at terapascal pressures, and \emph{sh}, \emph{fcc}, \emph{dhcp} and \emph{bcc} up to petapascal pressures.\cite{martinez2012thermodynamically} Carbon structures that are metastable under all conditions include graphene, nanotubes, and fullerenes.\cite{Lu2022}

We will now explore what can be learned about carbon from a single known carbon phase - the diamond structure. This high symmetry Fd$\bar{3}$m cubic structure has a single environment, so the generated structures will be optimised to have environments as close to this environment as possible. 

A cost-based EDDP potential was generated using the \code{manifest} script which performs the active learning process. A three-body neural network potential with 16 polynomials for the two-body terms of the environment features, and 4 for the three-body was trained, with two hidden layers of 20 nodes each. 31 individual networks were trained, with 18 selected by the NNLS ensembling procedure. 1000 structures with 1 to 12 atoms were randomly generated in the first step, along with 1000 shakes of the target diamond structure with a position and cell amplitude of 0.1. The cutoff radius was set to 3.75 \AA. During the active learning phase 10 cycles of adding 1000 AIRSS generated structures, added with a 0.1 position and cell amplitude shake. Parameters for the cost function were $\alpha=10$ and $\beta=100$.

A search for low energy carbon structures was performed in the following way. Using the cost-based EDDP an AIRSS search is conducted for 8 to 48 atoms, generating initial structures with a volume per atom between 5 and 10 \AA$^3$ and 12 to 24 randomly selected symmetry operations. The application of high symmetry ensures a diversity of generated structures, and at the same time reduces the number of low energy structures which are simply defected versions of diamond or graphite. The ranking of the structures is performed in three stages, using PBE-DFT,\cite{perdew1996generalized} computed by CASTEP.\cite{clark2005first} First, single point DFT energies are computed for all the generated structures using the following settings: the default QC5 OTFG pseudopotential for carbon, a k-point spacing of 0.07 $2\pi$ \AA$^{-1}$, plane wave cutoff of 340 eV, and default grid scales. Next, all structures within 1 eV of the most stable structure are DFT geometry optimised with the same settings. Finally the structures within 0.5 eV of the ground state are re-optimised with more stringent settings: the default C9 OTFG pseudopotential for carbon, a k-point spacing of 0.03 $2\pi$ \AA$^{-1}$, plane wave cutoff of 700 eV, with standard and fine grid scales of 2 and 2.3 respectively.

\begin{figure}
\includegraphics[width=0.45\textwidth]{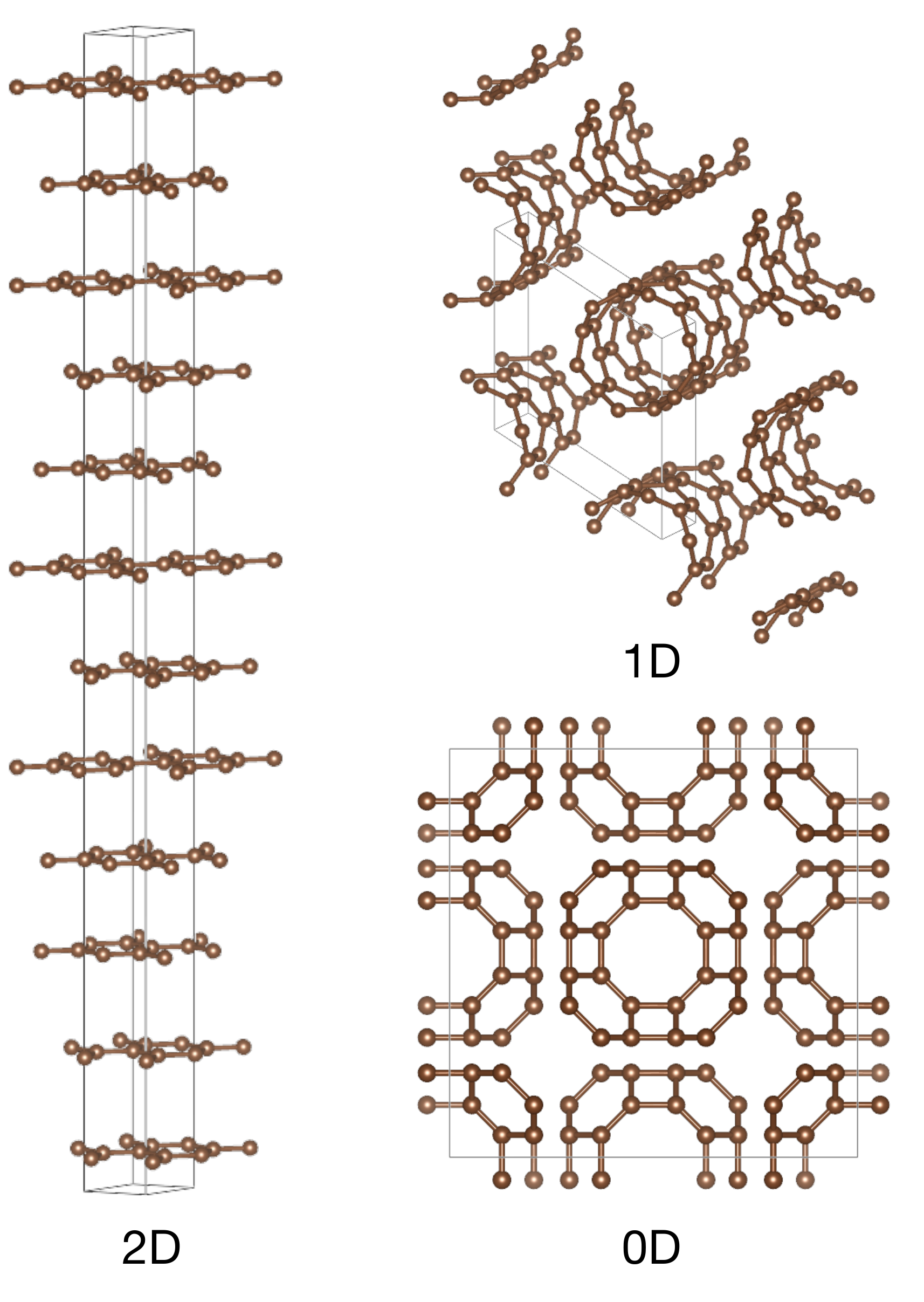}
\caption{\emph{Selected low dimensional carbon structures} The zero-dimensional structure consists of a face-centered lattice of C$_{48}$ clusters, but relatively unstable compared to the fullerenes due to the presence of four membered rings. The one-dimensional structure is an array of small nanotubes, and the two-dimensional structure is a complex stacking of graphite.}
\label{carbon-low}
\end{figure}

Analysing the structures up to 1 eV reveal a wide variety of bonding beyond that of the tetrahedral diamond from which the structures are generated, including $sp$, $sp^2$ and $sp^3$ bonding and mixtures. In Figure \ref{carbon-low} the most stable zero, one and two dimensional structures are highlighted. The observation that, starting from the experimental diamond structure, isolated clusters (foreshadowing the fullerenes), nanotubes and graphitic structures are generated is astonishing, and suggests the discovery potential of single pieces of data. Even without the DFT energetic data, which points to a given structure's stability and likely synthesisabilty, the existence of the low dimensional, threefold coordinated, carbon structures among the generated structures would likely encourage speculation, had they not been previously known. It should be noted that the application of symmetry enforces the large diversity of structures. However, even without applying symmetry in a search of 8 carbon atoms, layered graphitic like structures are generated, albeit somewhat distorted, and highly compressed.

\begin{figure}
\includegraphics[width=0.45\textwidth]{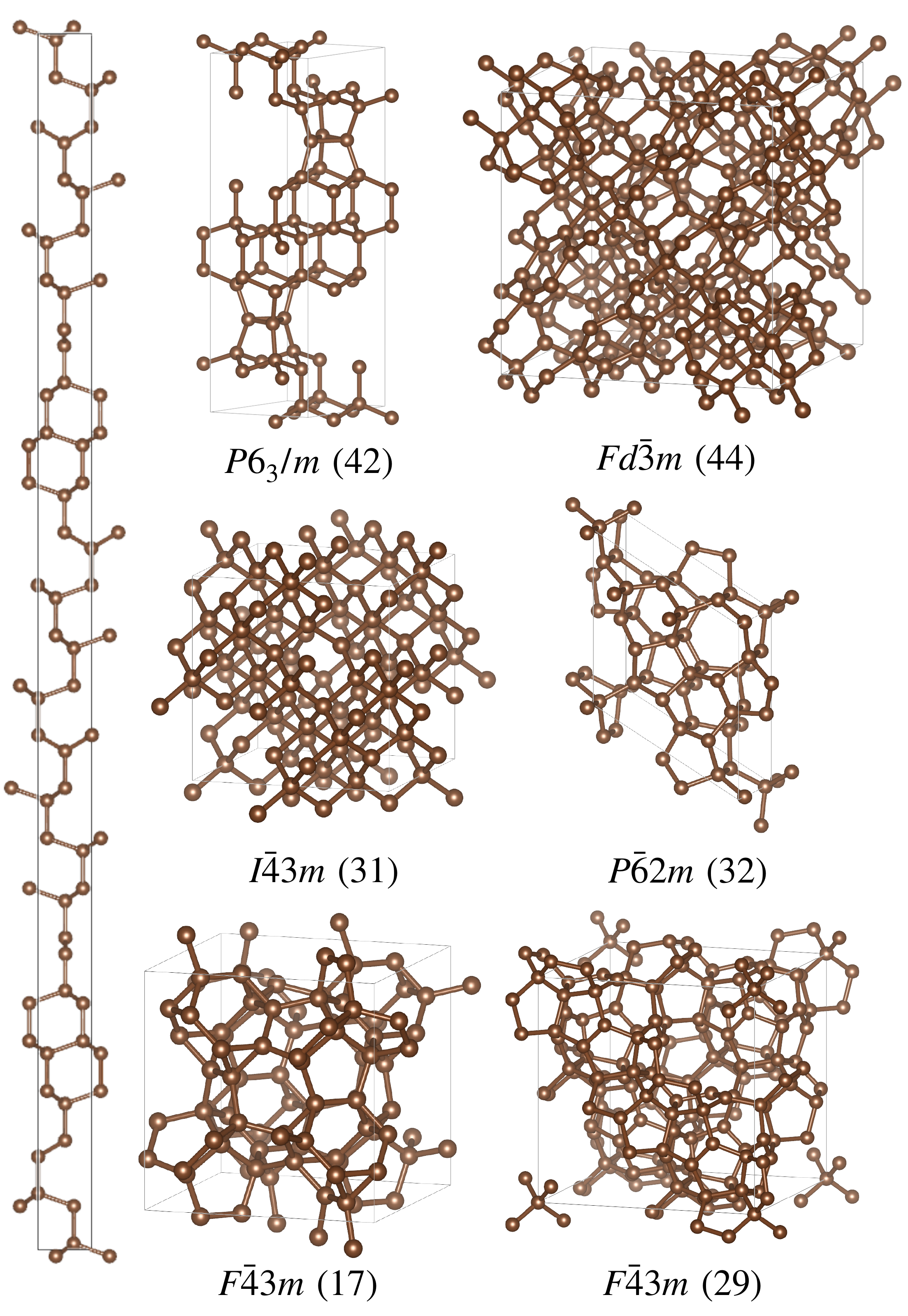}
\caption{\emph{Selected three-dimensional carbon framework structures} The space groups and number of atoms in the primitive unit cell are indicated. The left hand, high aspect ratio, structure has space group $P6_122$ and 48 atoms. It is characterised by regions of diamond-like material, connected by graphitic regions, reminiscent of \emph{diaphite}.\cite{nemeth2021diaphite}}
\label{carbon}
\end{figure}

The data relaxed to a higher level of accuracy up to 0.5 eV above the most stable structures are filtered so as to highlight only the three-dimensional carbon framework structures. The resulting structures are listed in Table \ref{carbon-list} and a selection highlighted in Figure \ref{carbon}. The SACADA\cite{hoffmann2016homo} online database aims to collect the many, often repeated, predictions of carbon structures from the literature. This is a challenging task, and absence in the database does not necessarily indicate the novelty of a given structure. Further, many topologies may have been reported for related systems such as silicon, and the silicates. However, it is notable that a significant fraction of the structures reported in Table \ref{carbon-list} are not currently listed in the SACADA database, again pointing to the discovery potential of generating structures related to a single known experimental structure.

\begin{table}[]
\begin{tabular}{lcccc}
Space Group & Number & Energy (eV) & Volume (\AA$^3$) & SACADA \# \\
\hline
 Fd$\bar3$m & 34 & 0.205 & 6.583 & 158 \\
 Pm$\bar3$n & 46 & 0.238 & 6.526 & 159 \\
 P4$_2$/ncm & 12 & 0.239 & 5.954 & 107 \\
 Pn$\bar3$m & 24 & 0.241 & 9.411 & 46 \\
 P6$_5$22 & 6 & 0.242 & 6.213 & 29 \\
 P6$_3$/mcm & 48 & 0.260 & 5.855 & 917 \\
 P6$_3$/mmc & 36 & 0.292 & 5.908 & 549 \\
 P6$_3$/m & 42 & 0.316 & 5.948 & - \\
 P6$_1$22 & 36 & 0.323 & 5.907 & 569 \\
 I4/mmm & 4 & 0.328 & 6.011 & 60 \\
 Fd$\bar3$m & 44 & 0.341 & 7.029 & - \\
 I$\bar4$3m & 31 & 0.342 & 5.894 & - \\
 I$\bar4$3m & 23 & 0.346 & 6.293 & 204 \\
 P$\bar6$2m & 32 & 0.352 & 5.988 & - \\
 P6$_1$22 & 48 & 0.359 & 5.861 & - \\
 F$\bar4$3m & 17 & 0.365 & 7.223 & - \\
 P6$_1$22 & 48 & 0.373 & 5.868 & - \\
 P$\bar6$2m & 15 & 0.378 & 6.043 & - \\
 P6$_3$/m & 48 & 0.380 & 6.427 & - \\
 I4/mmm & 16 & 0.383 & 5.848 & 916 \\
 P6$_3$22 & 48 & 0.386 & 6.048 & - \\
 P6/m & 48 & 0.388 & 6.048 & - \\
 P6/mmm & 12 & 0.427 & 6.049 & - \\
 I4$_1$/acd & 32 & 0.428 & 6.114 & - \\
 P$\bar3$c1 & 48 & 0.431 & 6.131 & - \\
 P6$_5$22 & 36 & 0.433 & 5.920 & - \\
 I4/mcm & 8 & 0.435 & 6.392 & 76 \\
 F$\bar4$3m & 29 & 0.436 & 7.904 & - \\
 P6/m & 16 & 0.436 & 7.611 & - \\
 P6/mmm & 36 & 0.455 & 6.193 & 1037 \\
 Im$\bar3$m & 24 & 0.462 & 10.046 & 54 \\
 P6/m & 34 & 0.475 & 6.328 & - \\
 Im$\bar3$m & 30 & 0.485 & 6.180 & 121 \\
 P4/mnc & 40 & 0.494 & 6.103 & - \\
\end{tabular}
\caption{\emph{Three-dimensional carbon framework structures} Space groups are reported in the Hermann–Mauguin notation, along with the number of atoms in the primitive unit cell. The total energies, with respect to the graphitic two dimensional structure shown in Figure \ref{carbon-low}, and volumes are reported per atom. The SACADA serial number is reported where identified. A dash indicates no SACADA entry has been identified.}
\label{carbon-list}
\end{table}

\subsection{Pyrope garnet Mg$_3$Al$_2$(SiO$_4$)$_3$}

To extend the investigation to a more complex example we consider the pyrope garnet composition, Mg$_3$Al$_2$(SiO$_4$)$_3$. The garnet structure is rather elaborate, Ia$\bar{3}$d cubic with 160 atoms in the conventional unit cell. With four chemical species, in contrast to the diamond structure there are multiple local environments.

\begin{figure}
\includegraphics[width=0.45\textwidth]{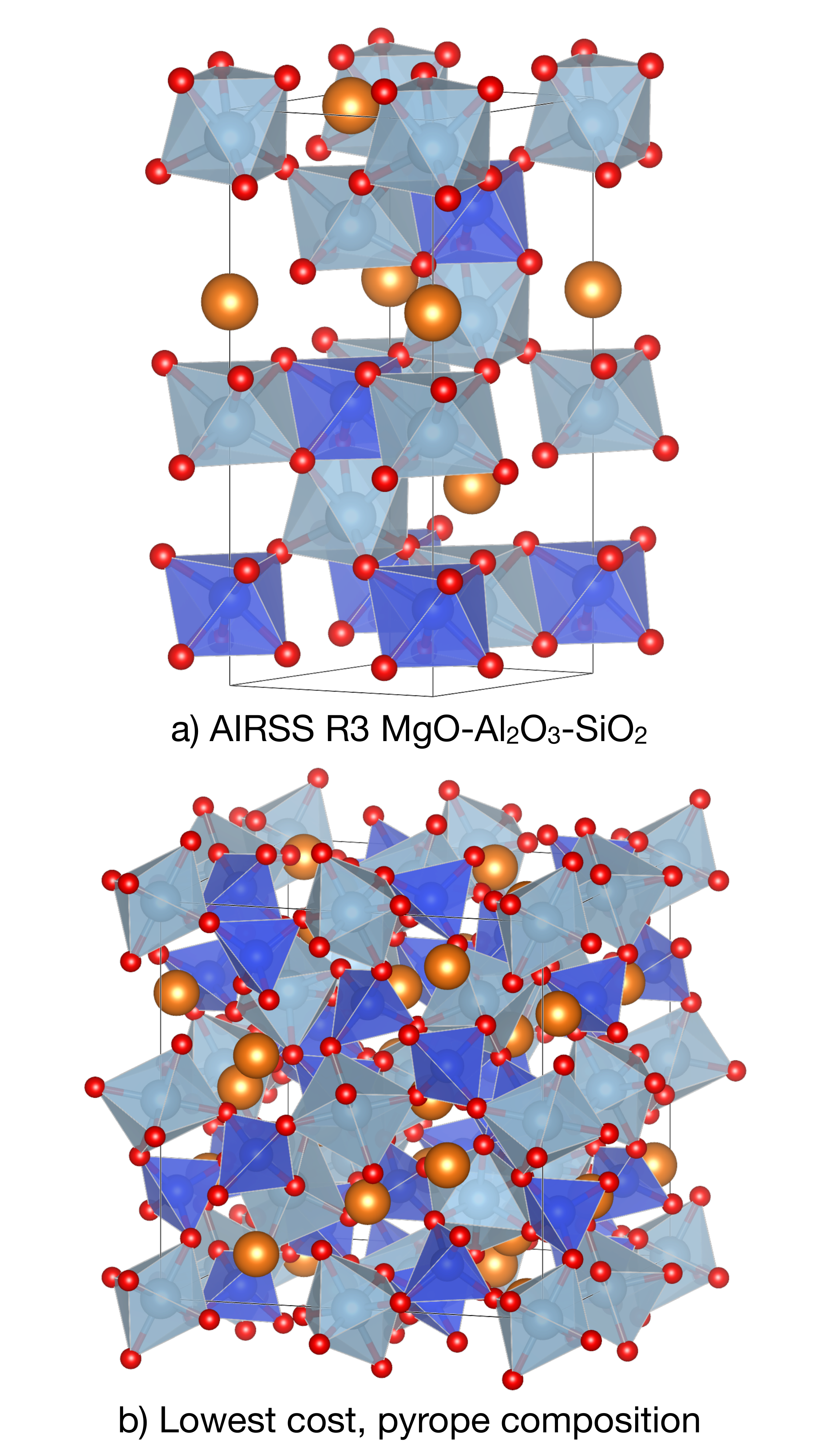}
\caption{\emph{Generation of pyrope garnet structure} a) The conventional cell of the R$3$ symmetry AIRSS generated structure for a single formula unit of MgO-Al$_2$O$_3$-SiO$_2$ at 10GPa. b) The lowest predicted cost structure in the pyrope Mg$_3$Al$_2$(SiO$_4$)$_3$ composition, which is identical to the experimentally known 180 atom conventional cell garnet structure.}
\label{pyrope}
\end{figure}

To explore the transferability of the approach, and to test its integration into a measurement-based structure searching strategy, rather than starting from the pyrope composition, or an experimental crystal structure, a DFT driven AIRSS search with a single formula unit of a 1:1:1 composition of MgO, Al$_2$O$_3$ and SiO$_2$ was first performed. The initial random structure were generated to have a range of volumes and a random MINSEP matrix of between 2 and 3 \AA. Symmetry was applied to the structures, randomly choosing 2 to 4 symmetry operations. CASTEP, QC5 OTFG pseudopotentals, a 340~eV plane wave cutoff, and 0.07 $2\pi$ \AA$^{-1}$ k-point spacing and the PBE density functional were used to structural optimise 29 random structures under 10~GPa of applied external pressure. A structure with the space group R$3$, see Figure \ref{pyrope},  was encountered multiple (6) times, and taken as the target structure for the generation of a cost-based EDDP potential.

A two-body EDDP was trained on the cost data using \code{manifest}, with 16 polynomials for the environment features, and two hidden layers of 20 nodes each. 30 individual networks were trained, with 9 selected by the NNLS ensembling procedure. 1000 structures with a single formula unit of MgO-Al$_2$O$_3$-SiO$_2$ were randomly generated in the first step, applying 2 to 4 symmetry operations and a random MINSEP matrix of 2 to 3 \AA, along with 1000 shakes of the target lowest energy MgO-Al$_2$O$_3$-SiO$_2$ structure with a position and cell amplitude of 0.1. The cutoff radius was set to 5 \AA. During the active learning phase there were 5 cycles of 1000 AIRSS generated structures, added with a 0.1 position and cell amplitude shake. Parameters for the cost function were $\alpha=100$ and $\beta=10$.

Using the cost-based EDDP a random search is performed in the pyrope, Mg$_3$Al$_2$(SiO$_4$)$_3$, composition, and a unit cell containing 4 formula units, 24 and 48 randomly chosen symmetry operations, and a random MINSEP matrix of 2 to 3 \AA. Of the 814 structures generated, the one with the lowest EDDP predicted cost had a space group of Ia$\bar{3}$d and was encountered three times. Already visually appearing very similar, geometry optimising the generated structure using CASTEP, QC5 OTFG pseudopotentials, a 340 eV plane wave cutoff and a gamma point sampling of the Brillouin Zone, leads to an identical structure to the experimentally known pyrope garnet. The next lowest predicted cost structure, with space group I$4_132$, is 223 meV/atom less stable when optimised at 10 GPa. The rediscovery of the garnet structure demonstrates both the transferability of the approach to novel compositions, and a practical and highly computationally efficient method to uncover complex crystal structures.

\section{Relation to diffusion based generative approaches}
\label{discuss}

It can be a challenge to navigate the differences in terminology when research fields collide. Generative machine learning methods have excited the research community. The field of structure prediction is no exception, with a wide array of generative approaches to structure prediction being explored.\cite{kim2020generative,court20203,pakornchote2024diffusion,luo2024deep,zeni2023mattergen,cheng2024response,ronne2024generative} In the above I have tried to make the case that the building of “random sensible structures” is a generative process. But the similarities to machine learning based approaches go beyond that. 

The scheme outlined in Section \ref{vectors} is in essence identical to a generative diffusion process.  In a diffusion model target images, or structures, are “noised” - or in the language of random structure searching “shaken”. The noise is increased until no remnants of the original target remains. Given the target, and the noised intermediates, a machine learning model is trained to “find its way” from a noised to a less noised configuration. As described illuminatingly in Ref. \onlinecite{permenter2023interpreting} the denoising can be achieved by starting from a random configuration and minimising some cost function of the distance to the manifold of the target examples. It is clear that this is exactly the procedure described in Section \ref{vectors}, where the machine learning model is an EDDP, trained on distance (in feature, or environment vector, space) derived data. Indeed, it is clear that such a diffusion style model is also very similar to random structure search based on an EDDP (or other MLIP) trained on DFT energetic data of marker structures - and going  downhill in energy takes you back to the marker structures, or new similar ones, with similarly low energy.  From this perspective it is instructive to note the fundamental similarity of generative models (such as MatterGen\cite{zeni2023mattergen}), and universal potentials (such as MACE0\cite{batatia2023foundation}) coupled with AIRSS.\cite{pickard2006high,pickard2011ab}

When creating diffusion models, a lot of care is taken in designing the noising process. From the perspective of structure prediction, this is equivalent to designing appropriate shakes in AIRSS, or moves in basin hopping style algorithms. This suggests that there is expected to be considerable benefit from exploring the respective field's insights - for the generative models to learning the denoising process, and for MLIPs to design optimal sampling of energy landscapes for the construction of training datasets.

\section{Conclusion}

First principles random structure searching has proven to be an engine for the discovery of novel arrangements of matter, exposing new science - almost to the point of being routine. With the rise of data driven methods - especially the machine learned interatomic potentials, but also the closely related generative approaches, AIRSS is emerging as a one of the most important sources of the data itself. With innovations enabled by machine learning acceleration, such as hot-AIRSS, which broadens the applicability of AIRSS to greater numbers of ever more complex structures, and more sophisticated schemes for generating candidate structures, data driven discovery is emerging as a powerful force in the atomistic sciences.

\bibliography{fd}

\begin{thebibliography}{81}%
\makeatletter
\providecommand \@ifxundefined [1]{%
 \@ifx{#1\undefined}
}%
\providecommand \@ifnum [1]{%
 \ifnum #1\expandafter \@firstoftwo
 \else \expandafter \@secondoftwo
 \fi
}%
\providecommand \@ifx [1]{%
 \ifx #1\expandafter \@firstoftwo
 \else \expandafter \@secondoftwo
 \fi
}%
\providecommand \natexlab [1]{#1}%
\providecommand \enquote  [1]{``#1''}%
\providecommand \bibnamefont  [1]{#1}%
\providecommand \bibfnamefont [1]{#1}%
\providecommand \citenamefont [1]{#1}%
\providecommand \href@noop [0]{\@secondoftwo}%
\providecommand \href [0]{\begingroup \@sanitize@url \@href}%
\providecommand \@href[1]{\@@startlink{#1}\@@href}%
\providecommand \@@href[1]{\endgroup#1\@@endlink}%
\providecommand \@sanitize@url [0]{\catcode `\\12\catcode `\$12\catcode
  `\&12\catcode `\#12\catcode `\^12\catcode `\_12\catcode `\%12\relax}%
\providecommand \@@startlink[1]{}%
\providecommand \@@endlink[0]{}%
\providecommand \url  [0]{\begingroup\@sanitize@url \@url }%
\providecommand \@url [1]{\endgroup\@href {#1}{\urlprefix }}%
\providecommand \urlprefix  [0]{URL }%
\providecommand \Eprint [0]{\href }%
\providecommand \doibase [0]{http://dx.doi.org/}%
\providecommand \selectlanguage [0]{\@gobble}%
\providecommand \bibinfo  [0]{\@secondoftwo}%
\providecommand \bibfield  [0]{\@secondoftwo}%
\providecommand \translation [1]{[#1]}%
\providecommand \BibitemOpen [0]{}%
\providecommand \bibitemStop [0]{}%
\providecommand \bibitemNoStop [0]{.\EOS\space}%
\providecommand \EOS [0]{\spacefactor3000\relax}%
\providecommand \BibitemShut  [1]{\csname bibitem#1\endcsname}%
\let\auto@bib@innerbib\@empty
\bibitem [{\citenamefont {Oganov}\ \emph {et~al.}(2019)\citenamefont {Oganov},
  \citenamefont {Pickard}, \citenamefont {Zhu},\ and\ \citenamefont
  {Needs}}]{oganov2019structure}%
  \BibitemOpen
  \bibfield  {author} {\bibinfo {author} {\bibfnamefont {A.~R.}\ \bibnamefont
  {Oganov}}, \bibinfo {author} {\bibfnamefont {C.~J.}\ \bibnamefont {Pickard}},
  \bibinfo {author} {\bibfnamefont {Q.}~\bibnamefont {Zhu}}, \ and\ \bibinfo
  {author} {\bibfnamefont {R.~J.}\ \bibnamefont {Needs}},\ }\href@noop {}
  {\bibfield  {journal} {\bibinfo  {journal} {Nature Reviews Materials}\
  }\textbf {\bibinfo {volume} {4}},\ \bibinfo {pages} {331} (\bibinfo {year}
  {2019})}\BibitemShut {NoStop}%
\bibitem [{\citenamefont {Stillinger}\ and\ \citenamefont
  {Weber}(1985)}]{stillinger1985computer}%
  \BibitemOpen
  \bibfield  {author} {\bibinfo {author} {\bibfnamefont {F.~H.}\ \bibnamefont
  {Stillinger}}\ and\ \bibinfo {author} {\bibfnamefont {T.~A.}\ \bibnamefont
  {Weber}},\ }\href@noop {} {\bibfield  {journal} {\bibinfo  {journal}
  {Physical Review B}\ }\textbf {\bibinfo {volume} {31}},\ \bibinfo {pages}
  {5262} (\bibinfo {year} {1985})}\BibitemShut {NoStop}%
\bibitem [{\citenamefont {Biswas}\ and\ \citenamefont
  {Hamann}(1985)}]{biswas1985interatomic}%
  \BibitemOpen
  \bibfield  {author} {\bibinfo {author} {\bibfnamefont {R.}~\bibnamefont
  {Biswas}}\ and\ \bibinfo {author} {\bibfnamefont {D.}~\bibnamefont
  {Hamann}},\ }\href@noop {} {\bibfield  {journal} {\bibinfo  {journal}
  {Physical Review Letters}\ }\textbf {\bibinfo {volume} {55}},\ \bibinfo
  {pages} {2001} (\bibinfo {year} {1985})}\BibitemShut {NoStop}%
\bibitem [{\citenamefont {Tersoff}(1988)}]{tersoff1988empirical}%
  \BibitemOpen
  \bibfield  {author} {\bibinfo {author} {\bibfnamefont {J.}~\bibnamefont
  {Tersoff}},\ }\href@noop {} {\bibfield  {journal} {\bibinfo  {journal}
  {Physical Review Letters}\ }\textbf {\bibinfo {volume} {61}},\ \bibinfo
  {pages} {2879} (\bibinfo {year} {1988})}\BibitemShut {NoStop}%
\bibitem [{\citenamefont {Woodley}\ \emph {et~al.}(1999)\citenamefont
  {Woodley}, \citenamefont {Battle}, \citenamefont {Gale},\ and\ \citenamefont
  {Catlow}}]{woodley1999prediction}%
  \BibitemOpen
  \bibfield  {author} {\bibinfo {author} {\bibfnamefont {S.}~\bibnamefont
  {Woodley}}, \bibinfo {author} {\bibfnamefont {P.}~\bibnamefont {Battle}},
  \bibinfo {author} {\bibfnamefont {J.}~\bibnamefont {Gale}}, \ and\ \bibinfo
  {author} {\bibfnamefont {C.~A.}\ \bibnamefont {Catlow}},\ }\href@noop {}
  {\bibfield  {journal} {\bibinfo  {journal} {Physical Chemistry Chemical
  Physics}\ }\textbf {\bibinfo {volume} {1}},\ \bibinfo {pages} {2535}
  (\bibinfo {year} {1999})}\BibitemShut {NoStop}%
\bibitem [{\citenamefont {Lejaeghere}\ \emph {et~al.}(2016)\citenamefont
  {Lejaeghere}, \citenamefont {Bihlmayer}, \citenamefont {Bj{\"o}rkman},
  \citenamefont {Blaha}, \citenamefont {Bl{\"u}gel}, \citenamefont {Blum},
  \citenamefont {Caliste}, \citenamefont {Castelli}, \citenamefont {Clark},
  \citenamefont {Dal~Corso} \emph {et~al.}}]{lejaeghere2016reproducibility}%
  \BibitemOpen
  \bibfield  {author} {\bibinfo {author} {\bibfnamefont {K.}~\bibnamefont
  {Lejaeghere}}, \bibinfo {author} {\bibfnamefont {G.}~\bibnamefont
  {Bihlmayer}}, \bibinfo {author} {\bibfnamefont {T.}~\bibnamefont
  {Bj{\"o}rkman}}, \bibinfo {author} {\bibfnamefont {P.}~\bibnamefont {Blaha}},
  \bibinfo {author} {\bibfnamefont {S.}~\bibnamefont {Bl{\"u}gel}}, \bibinfo
  {author} {\bibfnamefont {V.}~\bibnamefont {Blum}}, \bibinfo {author}
  {\bibfnamefont {D.}~\bibnamefont {Caliste}}, \bibinfo {author} {\bibfnamefont
  {I.~E.}\ \bibnamefont {Castelli}}, \bibinfo {author} {\bibfnamefont {S.~J.}\
  \bibnamefont {Clark}}, \bibinfo {author} {\bibfnamefont {A.}~\bibnamefont
  {Dal~Corso}},  \emph {et~al.},\ }\href@noop {} {\bibfield  {journal}
  {\bibinfo  {journal} {Science}\ }\textbf {\bibinfo {volume} {351}},\ \bibinfo
  {pages} {aad3000} (\bibinfo {year} {2016})}\BibitemShut {NoStop}%
\bibitem [{\citenamefont {Brown}\ \emph {et~al.}(1996)\citenamefont {Brown},
  \citenamefont {Gibbs},\ and\ \citenamefont {Clary}}]{brown1996combining}%
  \BibitemOpen
  \bibfield  {author} {\bibinfo {author} {\bibfnamefont {D.~F.~R.}\
  \bibnamefont {Brown}}, \bibinfo {author} {\bibfnamefont {M.~N.}\ \bibnamefont
  {Gibbs}}, \ and\ \bibinfo {author} {\bibfnamefont {D.~C.}\ \bibnamefont
  {Clary}},\ }\href@noop {} {\bibfield  {journal} {\bibinfo  {journal} {The
  Journal of Chemical Physics}\ }\textbf {\bibinfo {volume} {105}},\ \bibinfo
  {pages} {7597} (\bibinfo {year} {1996})}\BibitemShut {NoStop}%
\bibitem [{\citenamefont {Behler}\ and\ \citenamefont
  {Parrinello}(2007)}]{behler2007generalized}%
  \BibitemOpen
  \bibfield  {author} {\bibinfo {author} {\bibfnamefont {J.}~\bibnamefont
  {Behler}}\ and\ \bibinfo {author} {\bibfnamefont {M.}~\bibnamefont
  {Parrinello}},\ }\href@noop {} {\bibfield  {journal} {\bibinfo  {journal}
  {Physical Review Letters}\ }\textbf {\bibinfo {volume} {98}},\ \bibinfo
  {pages} {146401} (\bibinfo {year} {2007})}\BibitemShut {NoStop}%
\bibitem [{\citenamefont {Bart{\'o}k}\ \emph {et~al.}(2010)\citenamefont
  {Bart{\'o}k}, \citenamefont {Payne}, \citenamefont {Kondor},\ and\
  \citenamefont {Cs{\'a}nyi}}]{bartok2010gaussian}%
  \BibitemOpen
  \bibfield  {author} {\bibinfo {author} {\bibfnamefont {A.~P.}\ \bibnamefont
  {Bart{\'o}k}}, \bibinfo {author} {\bibfnamefont {M.~C.}\ \bibnamefont
  {Payne}}, \bibinfo {author} {\bibfnamefont {R.}~\bibnamefont {Kondor}}, \
  and\ \bibinfo {author} {\bibfnamefont {G.}~\bibnamefont {Cs{\'a}nyi}},\
  }\href@noop {} {\bibfield  {journal} {\bibinfo  {journal} {Physical Review
  Letters}\ }\textbf {\bibinfo {volume} {104}},\ \bibinfo {pages} {136403}
  (\bibinfo {year} {2010})}\BibitemShut {NoStop}%
\bibitem [{\citenamefont {Bart{\'o}k}\ \emph {et~al.}(2017)\citenamefont
  {Bart{\'o}k}, \citenamefont {De}, \citenamefont {Poelking}, \citenamefont
  {Bernstein}, \citenamefont {Kermode}, \citenamefont {Cs{\'a}nyi},\ and\
  \citenamefont {Ceriotti}}]{bartok2017machine}%
  \BibitemOpen
  \bibfield  {author} {\bibinfo {author} {\bibfnamefont {A.~P.}\ \bibnamefont
  {Bart{\'o}k}}, \bibinfo {author} {\bibfnamefont {S.}~\bibnamefont {De}},
  \bibinfo {author} {\bibfnamefont {C.}~\bibnamefont {Poelking}}, \bibinfo
  {author} {\bibfnamefont {N.}~\bibnamefont {Bernstein}}, \bibinfo {author}
  {\bibfnamefont {J.~R.}\ \bibnamefont {Kermode}}, \bibinfo {author}
  {\bibfnamefont {G.}~\bibnamefont {Cs{\'a}nyi}}, \ and\ \bibinfo {author}
  {\bibfnamefont {M.}~\bibnamefont {Ceriotti}},\ }\href@noop {} {\bibfield
  {journal} {\bibinfo  {journal} {Science Advances}\ }\textbf {\bibinfo
  {volume} {3}},\ \bibinfo {pages} {e1701816} (\bibinfo {year}
  {2017})}\BibitemShut {NoStop}%
\bibitem [{\citenamefont {Pickard}(2022)}]{pickard2022ephemeral}%
  \BibitemOpen
  \bibfield  {author} {\bibinfo {author} {\bibfnamefont {C.~J.}\ \bibnamefont
  {Pickard}},\ }\href@noop {} {\bibfield  {journal} {\bibinfo  {journal}
  {Physical Review B}\ }\textbf {\bibinfo {volume} {106}},\ \bibinfo {pages}
  {014102} (\bibinfo {year} {2022})}\BibitemShut {NoStop}%
\bibitem [{\citenamefont {Salzbrenner}\ \emph {et~al.}(2023)\citenamefont
  {Salzbrenner}, \citenamefont {Joo}, \citenamefont {Conway}, \citenamefont
  {Cooke}, \citenamefont {Zhu}, \citenamefont {Matraszek}, \citenamefont
  {Witt},\ and\ \citenamefont {Pickard}}]{salzbrenner2023developments}%
  \BibitemOpen
  \bibfield  {author} {\bibinfo {author} {\bibfnamefont {P.~T.}\ \bibnamefont
  {Salzbrenner}}, \bibinfo {author} {\bibfnamefont {S.~H.}\ \bibnamefont
  {Joo}}, \bibinfo {author} {\bibfnamefont {L.~J.}\ \bibnamefont {Conway}},
  \bibinfo {author} {\bibfnamefont {P.~I.}\ \bibnamefont {Cooke}}, \bibinfo
  {author} {\bibfnamefont {B.}~\bibnamefont {Zhu}}, \bibinfo {author}
  {\bibfnamefont {M.~P.}\ \bibnamefont {Matraszek}}, \bibinfo {author}
  {\bibfnamefont {W.~C.}\ \bibnamefont {Witt}}, \ and\ \bibinfo {author}
  {\bibfnamefont {C.~J.}\ \bibnamefont {Pickard}},\ }\href@noop {} {\bibfield
  {journal} {\bibinfo  {journal} {The Journal of Chemical Physics}\ }\textbf
  {\bibinfo {volume} {159}} (\bibinfo {year} {2023})}\BibitemShut {NoStop}%
\bibitem [{\citenamefont {Pickard}\ and\ \citenamefont
  {Needs}(2006)}]{pickard2006high}%
  \BibitemOpen
  \bibfield  {author} {\bibinfo {author} {\bibfnamefont {C.~J.}\ \bibnamefont
  {Pickard}}\ and\ \bibinfo {author} {\bibfnamefont {R.~J.}\ \bibnamefont
  {Needs}},\ }\href@noop {} {\bibfield  {journal} {\bibinfo  {journal}
  {Physical Review Letters}\ }\textbf {\bibinfo {volume} {97}},\ \bibinfo
  {pages} {045504} (\bibinfo {year} {2006})}\BibitemShut {NoStop}%
\bibitem [{\citenamefont {Pickard}\ and\ \citenamefont
  {Needs}(2011{\natexlab{a}})}]{pickard2011ab}%
  \BibitemOpen
  \bibfield  {author} {\bibinfo {author} {\bibfnamefont {C.~J.}\ \bibnamefont
  {Pickard}}\ and\ \bibinfo {author} {\bibfnamefont {R.~J.}\ \bibnamefont
  {Needs}},\ }\href@noop {} {\bibfield  {journal} {\bibinfo  {journal} {Journal
  of Physics: Condensed Matter}\ }\textbf {\bibinfo {volume} {23}},\ \bibinfo
  {pages} {053201} (\bibinfo {year} {2011}{\natexlab{a}})}\BibitemShut
  {NoStop}%
\bibitem [{\citenamefont {Pickard}\ and\ \citenamefont
  {Needs}(2007{\natexlab{a}})}]{pickard2007metallization}%
  \BibitemOpen
  \bibfield  {author} {\bibinfo {author} {\bibfnamefont {C.~J.}\ \bibnamefont
  {Pickard}}\ and\ \bibinfo {author} {\bibfnamefont {R.}~\bibnamefont
  {Needs}},\ }\href@noop {} {\bibfield  {journal} {\bibinfo  {journal}
  {Physical Review B}\ }\textbf {\bibinfo {volume} {76}},\ \bibinfo {pages}
  {144114} (\bibinfo {year} {2007}{\natexlab{a}})}\BibitemShut {NoStop}%
\bibitem [{\citenamefont {Pickard}\ \emph {et~al.}(2020)\citenamefont
  {Pickard}, \citenamefont {Errea},\ and\ \citenamefont
  {Eremets}}]{pickard2020superconducting}%
  \BibitemOpen
  \bibfield  {author} {\bibinfo {author} {\bibfnamefont {C.~J.}\ \bibnamefont
  {Pickard}}, \bibinfo {author} {\bibfnamefont {I.}~\bibnamefont {Errea}}, \
  and\ \bibinfo {author} {\bibfnamefont {M.~I.}\ \bibnamefont {Eremets}},\
  }\href@noop {} {\bibfield  {journal} {\bibinfo  {journal} {Annual Review of
  Condensed Matter Physics}\ }\textbf {\bibinfo {volume} {11}},\ \bibinfo
  {pages} {57} (\bibinfo {year} {2020})}\BibitemShut {NoStop}%
\bibitem [{\citenamefont {Oganov}\ and\ \citenamefont
  {Glass}(2006)}]{oganov2006crystal}%
  \BibitemOpen
  \bibfield  {author} {\bibinfo {author} {\bibfnamefont {A.~R.}\ \bibnamefont
  {Oganov}}\ and\ \bibinfo {author} {\bibfnamefont {C.~W.}\ \bibnamefont
  {Glass}},\ }\href@noop {} {\bibfield  {journal} {\bibinfo  {journal} {The
  Journal of Chemical Physics}\ }\textbf {\bibinfo {volume} {124}},\ \bibinfo
  {pages} {244704} (\bibinfo {year} {2006})}\BibitemShut {NoStop}%
\bibitem [{\citenamefont {Wang}\ \emph {et~al.}(2012)\citenamefont {Wang},
  \citenamefont {Lv}, \citenamefont {Zhu},\ and\ \citenamefont
  {Ma}}]{wang2012calypso}%
  \BibitemOpen
  \bibfield  {author} {\bibinfo {author} {\bibfnamefont {Y.}~\bibnamefont
  {Wang}}, \bibinfo {author} {\bibfnamefont {J.}~\bibnamefont {Lv}}, \bibinfo
  {author} {\bibfnamefont {L.}~\bibnamefont {Zhu}}, \ and\ \bibinfo {author}
  {\bibfnamefont {Y.}~\bibnamefont {Ma}},\ }\href@noop {} {\bibfield  {journal}
  {\bibinfo  {journal} {Computer Physics Communications}\ }\textbf {\bibinfo
  {volume} {183}},\ \bibinfo {pages} {2063} (\bibinfo {year}
  {2012})}\BibitemShut {NoStop}%
\bibitem [{\citenamefont {Lonie}\ and\ \citenamefont
  {Zurek}(2011)}]{lonie2011xtalopt}%
  \BibitemOpen
  \bibfield  {author} {\bibinfo {author} {\bibfnamefont {D.~C.}\ \bibnamefont
  {Lonie}}\ and\ \bibinfo {author} {\bibfnamefont {E.}~\bibnamefont {Zurek}},\
  }\href@noop {} {\bibfield  {journal} {\bibinfo  {journal} {Computer Physics
  Communications}\ }\textbf {\bibinfo {volume} {182}},\ \bibinfo {pages} {372}
  (\bibinfo {year} {2011})}\BibitemShut {NoStop}%
\bibitem [{\citenamefont {Lu}\ \emph {et~al.}(2021)\citenamefont {Lu},
  \citenamefont {Zhu}, \citenamefont {Shires}, \citenamefont {Scanlon},\ and\
  \citenamefont {Pickard}}]{lu2021ab}%
  \BibitemOpen
  \bibfield  {author} {\bibinfo {author} {\bibfnamefont {Z.}~\bibnamefont
  {Lu}}, \bibinfo {author} {\bibfnamefont {B.}~\bibnamefont {Zhu}}, \bibinfo
  {author} {\bibfnamefont {B.~W.}\ \bibnamefont {Shires}}, \bibinfo {author}
  {\bibfnamefont {D.~O.}\ \bibnamefont {Scanlon}}, \ and\ \bibinfo {author}
  {\bibfnamefont {C.~J.}\ \bibnamefont {Pickard}},\ }\href@noop {} {\bibfield
  {journal} {\bibinfo  {journal} {The Journal of Chemical Physics}\ }\textbf
  {\bibinfo {volume} {154}} (\bibinfo {year} {2021})}\BibitemShut {NoStop}%
\bibitem [{\citenamefont {Zhu}\ \emph {et~al.}(2021)\citenamefont {Zhu},
  \citenamefont {Lu}, \citenamefont {Pickard},\ and\ \citenamefont
  {Scanlon}}]{zhu2021accelerating}%
  \BibitemOpen
  \bibfield  {author} {\bibinfo {author} {\bibfnamefont {B.}~\bibnamefont
  {Zhu}}, \bibinfo {author} {\bibfnamefont {Z.}~\bibnamefont {Lu}}, \bibinfo
  {author} {\bibfnamefont {C.~J.}\ \bibnamefont {Pickard}}, \ and\ \bibinfo
  {author} {\bibfnamefont {D.~O.}\ \bibnamefont {Scanlon}},\ }\href@noop {}
  {\bibfield  {journal} {\bibinfo  {journal} {APL Materials}\ }\textbf
  {\bibinfo {volume} {9}} (\bibinfo {year} {2021})}\BibitemShut {NoStop}%
\bibitem [{\citenamefont {Smalley}\ \emph {et~al.}(2022)\citenamefont
  {Smalley}, \citenamefont {Hoskyns}, \citenamefont {Hughes}, \citenamefont
  {Johnstone}, \citenamefont {Willhammar}, \citenamefont {Young}, \citenamefont
  {Pickard}, \citenamefont {Logsdail}, \citenamefont {Midgley},\ and\
  \citenamefont {Harris}}]{smalley2022structure}%
  \BibitemOpen
  \bibfield  {author} {\bibinfo {author} {\bibfnamefont {C.~J.}\ \bibnamefont
  {Smalley}}, \bibinfo {author} {\bibfnamefont {H.~E.}\ \bibnamefont
  {Hoskyns}}, \bibinfo {author} {\bibfnamefont {C.~E.}\ \bibnamefont {Hughes}},
  \bibinfo {author} {\bibfnamefont {D.~N.}\ \bibnamefont {Johnstone}}, \bibinfo
  {author} {\bibfnamefont {T.}~\bibnamefont {Willhammar}}, \bibinfo {author}
  {\bibfnamefont {M.~T.}\ \bibnamefont {Young}}, \bibinfo {author}
  {\bibfnamefont {C.~J.}\ \bibnamefont {Pickard}}, \bibinfo {author}
  {\bibfnamefont {A.~J.}\ \bibnamefont {Logsdail}}, \bibinfo {author}
  {\bibfnamefont {P.~A.}\ \bibnamefont {Midgley}}, \ and\ \bibinfo {author}
  {\bibfnamefont {K.~D.}\ \bibnamefont {Harris}},\ }\href@noop {} {\bibfield
  {journal} {\bibinfo  {journal} {Chemical Science}\ }\textbf {\bibinfo
  {volume} {13}},\ \bibinfo {pages} {5277} (\bibinfo {year}
  {2022})}\BibitemShut {NoStop}%
\bibitem [{\citenamefont {Kapil}\ \emph {et~al.}(2022)\citenamefont {Kapil},
  \citenamefont {Schran}, \citenamefont {Zen}, \citenamefont {Chen},
  \citenamefont {Pickard},\ and\ \citenamefont {Michaelides}}]{kapil2022first}%
  \BibitemOpen
  \bibfield  {author} {\bibinfo {author} {\bibfnamefont {V.}~\bibnamefont
  {Kapil}}, \bibinfo {author} {\bibfnamefont {C.}~\bibnamefont {Schran}},
  \bibinfo {author} {\bibfnamefont {A.}~\bibnamefont {Zen}}, \bibinfo {author}
  {\bibfnamefont {J.}~\bibnamefont {Chen}}, \bibinfo {author} {\bibfnamefont
  {C.~J.}\ \bibnamefont {Pickard}}, \ and\ \bibinfo {author} {\bibfnamefont
  {A.}~\bibnamefont {Michaelides}},\ }\href@noop {} {\bibfield  {journal}
  {\bibinfo  {journal} {Nature}\ }\textbf {\bibinfo {volume} {609}},\ \bibinfo
  {pages} {512} (\bibinfo {year} {2022})}\BibitemShut {NoStop}%
\bibitem [{\citenamefont {Deringer}\ \emph {et~al.}(2018)\citenamefont
  {Deringer}, \citenamefont {Pickard},\ and\ \citenamefont
  {Cs{\'a}nyi}}]{deringer2018data}%
  \BibitemOpen
  \bibfield  {author} {\bibinfo {author} {\bibfnamefont {V.~L.}\ \bibnamefont
  {Deringer}}, \bibinfo {author} {\bibfnamefont {C.~J.}\ \bibnamefont
  {Pickard}}, \ and\ \bibinfo {author} {\bibfnamefont {G.}~\bibnamefont
  {Cs{\'a}nyi}},\ }\href@noop {} {\bibfield  {journal} {\bibinfo  {journal}
  {Physical Review Letters}\ }\textbf {\bibinfo {volume} {120}},\ \bibinfo
  {pages} {156001} (\bibinfo {year} {2018})}\BibitemShut {NoStop}%
\bibitem [{\citenamefont {Merchant}\ \emph {et~al.}(2023)\citenamefont
  {Merchant}, \citenamefont {Batzner}, \citenamefont {Schoenholz},
  \citenamefont {Aykol}, \citenamefont {Cheon},\ and\ \citenamefont
  {Cubuk}}]{merchant2023scaling}%
  \BibitemOpen
  \bibfield  {author} {\bibinfo {author} {\bibfnamefont {A.}~\bibnamefont
  {Merchant}}, \bibinfo {author} {\bibfnamefont {S.}~\bibnamefont {Batzner}},
  \bibinfo {author} {\bibfnamefont {S.~S.}\ \bibnamefont {Schoenholz}},
  \bibinfo {author} {\bibfnamefont {M.}~\bibnamefont {Aykol}}, \bibinfo
  {author} {\bibfnamefont {G.}~\bibnamefont {Cheon}}, \ and\ \bibinfo {author}
  {\bibfnamefont {E.~D.}\ \bibnamefont {Cubuk}},\ }\href@noop {} {\bibfield
  {journal} {\bibinfo  {journal} {Nature}\ }\textbf {\bibinfo {volume} {624}},\
  \bibinfo {pages} {80} (\bibinfo {year} {2023})}\BibitemShut {NoStop}%
\bibitem [{\citenamefont {Zeni}\ \emph {et~al.}(2023)\citenamefont {Zeni},
  \citenamefont {Pinsler}, \citenamefont {Z{\"u}gner}, \citenamefont {Fowler},
  \citenamefont {Horton}, \citenamefont {Fu}, \citenamefont {Shysheya},
  \citenamefont {Crabb{\'e}}, \citenamefont {Sun}, \citenamefont {Smith} \emph
  {et~al.}}]{zeni2023mattergen}%
  \BibitemOpen
  \bibfield  {author} {\bibinfo {author} {\bibfnamefont {C.}~\bibnamefont
  {Zeni}}, \bibinfo {author} {\bibfnamefont {R.}~\bibnamefont {Pinsler}},
  \bibinfo {author} {\bibfnamefont {D.}~\bibnamefont {Z{\"u}gner}}, \bibinfo
  {author} {\bibfnamefont {A.}~\bibnamefont {Fowler}}, \bibinfo {author}
  {\bibfnamefont {M.}~\bibnamefont {Horton}}, \bibinfo {author} {\bibfnamefont
  {X.}~\bibnamefont {Fu}}, \bibinfo {author} {\bibfnamefont {S.}~\bibnamefont
  {Shysheya}}, \bibinfo {author} {\bibfnamefont {J.}~\bibnamefont
  {Crabb{\'e}}}, \bibinfo {author} {\bibfnamefont {L.}~\bibnamefont {Sun}},
  \bibinfo {author} {\bibfnamefont {J.}~\bibnamefont {Smith}},  \emph
  {et~al.},\ }\href@noop {} {\bibfield  {journal} {\bibinfo  {journal} {arXiv
  preprint arXiv:2312.03687}\ } (\bibinfo {year} {2023})}\BibitemShut {NoStop}%
\bibitem [{\citenamefont {Pickard}\ and\ \citenamefont
  {Needs}(2007{\natexlab{b}})}]{pickard2007structure}%
  \BibitemOpen
  \bibfield  {author} {\bibinfo {author} {\bibfnamefont {C.~J.}\ \bibnamefont
  {Pickard}}\ and\ \bibinfo {author} {\bibfnamefont {R.~J.}\ \bibnamefont
  {Needs}},\ }\href@noop {} {\bibfield  {journal} {\bibinfo  {journal} {Nature
  Physics}\ }\textbf {\bibinfo {volume} {3}},\ \bibinfo {pages} {473} (\bibinfo
  {year} {2007}{\natexlab{b}})}\BibitemShut {NoStop}%
\bibitem [{\citenamefont {Loubeyre}\ \emph {et~al.}(2020)\citenamefont
  {Loubeyre}, \citenamefont {Occelli},\ and\ \citenamefont
  {Dumas}}]{loubeyre2020synchrotron}%
  \BibitemOpen
  \bibfield  {author} {\bibinfo {author} {\bibfnamefont {P.}~\bibnamefont
  {Loubeyre}}, \bibinfo {author} {\bibfnamefont {F.}~\bibnamefont {Occelli}}, \
  and\ \bibinfo {author} {\bibfnamefont {P.}~\bibnamefont {Dumas}},\
  }\href@noop {} {\bibfield  {journal} {\bibinfo  {journal} {Nature}\ }\textbf
  {\bibinfo {volume} {577}},\ \bibinfo {pages} {631} (\bibinfo {year}
  {2020})}\BibitemShut {NoStop}%
\bibitem [{\citenamefont {Monacelli}\ \emph {et~al.}(2023)\citenamefont
  {Monacelli}, \citenamefont {Casula}, \citenamefont {Nakano}, \citenamefont
  {Sorella},\ and\ \citenamefont {Mauri}}]{monacelli2023quantum}%
  \BibitemOpen
  \bibfield  {author} {\bibinfo {author} {\bibfnamefont {L.}~\bibnamefont
  {Monacelli}}, \bibinfo {author} {\bibfnamefont {M.}~\bibnamefont {Casula}},
  \bibinfo {author} {\bibfnamefont {K.}~\bibnamefont {Nakano}}, \bibinfo
  {author} {\bibfnamefont {S.}~\bibnamefont {Sorella}}, \ and\ \bibinfo
  {author} {\bibfnamefont {F.}~\bibnamefont {Mauri}},\ }\href@noop {}
  {\bibfield  {journal} {\bibinfo  {journal} {Nature Physics}\ }\textbf
  {\bibinfo {volume} {19}},\ \bibinfo {pages} {845} (\bibinfo {year}
  {2023})}\BibitemShut {NoStop}%
\bibitem [{\citenamefont {Howie}\ \emph {et~al.}(2012)\citenamefont {Howie},
  \citenamefont {Guillaume}, \citenamefont {Scheler}, \citenamefont
  {Goncharov},\ and\ \citenamefont {Gregoryanz}}]{howie2012mixed}%
  \BibitemOpen
  \bibfield  {author} {\bibinfo {author} {\bibfnamefont {R.~T.}\ \bibnamefont
  {Howie}}, \bibinfo {author} {\bibfnamefont {C.~L.}\ \bibnamefont
  {Guillaume}}, \bibinfo {author} {\bibfnamefont {T.}~\bibnamefont {Scheler}},
  \bibinfo {author} {\bibfnamefont {A.~F.}\ \bibnamefont {Goncharov}}, \ and\
  \bibinfo {author} {\bibfnamefont {E.}~\bibnamefont {Gregoryanz}},\
  }\href@noop {} {\bibfield  {journal} {\bibinfo  {journal} {Physical Review
  Letters}\ }\textbf {\bibinfo {volume} {108}},\ \bibinfo {pages} {125501}
  (\bibinfo {year} {2012})}\BibitemShut {NoStop}%
\bibitem [{\citenamefont {Price}(2014)}]{price2014predicting}%
  \BibitemOpen
  \bibfield  {author} {\bibinfo {author} {\bibfnamefont {S.~L.}\ \bibnamefont
  {Price}},\ }\href@noop {} {\bibfield  {journal} {\bibinfo  {journal}
  {Chemical Society Reviews}\ }\textbf {\bibinfo {volume} {43}},\ \bibinfo
  {pages} {2098} (\bibinfo {year} {2014})}\BibitemShut {NoStop}%
\bibitem [{\citenamefont {Pickard}\ and\ \citenamefont
  {Needs}(2008)}]{pickard2008highly}%
  \BibitemOpen
  \bibfield  {author} {\bibinfo {author} {\bibfnamefont {C.~J.}\ \bibnamefont
  {Pickard}}\ and\ \bibinfo {author} {\bibfnamefont {R.}~\bibnamefont
  {Needs}},\ }\href@noop {} {\bibfield  {journal} {\bibinfo  {journal} {Nature
  Materials}\ }\textbf {\bibinfo {volume} {7}},\ \bibinfo {pages} {775}
  (\bibinfo {year} {2008})}\BibitemShut {NoStop}%
\bibitem [{\citenamefont {Ninet}\ \emph {et~al.}(2014)\citenamefont {Ninet},
  \citenamefont {Datchi}, \citenamefont {Dumas}, \citenamefont {Mezouar},
  \citenamefont {Garbarino}, \citenamefont {Mafety}, \citenamefont {Pickard},
  \citenamefont {Needs},\ and\ \citenamefont {Saitta}}]{ninet2014experimental}%
  \BibitemOpen
  \bibfield  {author} {\bibinfo {author} {\bibfnamefont {S.}~\bibnamefont
  {Ninet}}, \bibinfo {author} {\bibfnamefont {F.}~\bibnamefont {Datchi}},
  \bibinfo {author} {\bibfnamefont {P.}~\bibnamefont {Dumas}}, \bibinfo
  {author} {\bibfnamefont {M.}~\bibnamefont {Mezouar}}, \bibinfo {author}
  {\bibfnamefont {G.}~\bibnamefont {Garbarino}}, \bibinfo {author}
  {\bibfnamefont {A.}~\bibnamefont {Mafety}}, \bibinfo {author} {\bibfnamefont
  {C.}~\bibnamefont {Pickard}}, \bibinfo {author} {\bibfnamefont
  {R.}~\bibnamefont {Needs}}, \ and\ \bibinfo {author} {\bibfnamefont
  {A.}~\bibnamefont {Saitta}},\ }\href@noop {} {\bibfield  {journal} {\bibinfo
  {journal} {Physical Review B}\ }\textbf {\bibinfo {volume} {89}},\ \bibinfo
  {pages} {174103} (\bibinfo {year} {2014})}\BibitemShut {NoStop}%
\bibitem [{\citenamefont {Pickard}\ and\ \citenamefont
  {Needs}(2009)}]{pickard2009dense}%
  \BibitemOpen
  \bibfield  {author} {\bibinfo {author} {\bibfnamefont {C.~J.}\ \bibnamefont
  {Pickard}}\ and\ \bibinfo {author} {\bibfnamefont {R.}~\bibnamefont
  {Needs}},\ }\href@noop {} {\bibfield  {journal} {\bibinfo  {journal}
  {Physical Review Letters}\ }\textbf {\bibinfo {volume} {102}},\ \bibinfo
  {pages} {146401} (\bibinfo {year} {2009})}\BibitemShut {NoStop}%
\bibitem [{\citenamefont {Ma}\ \emph {et~al.}(2009)\citenamefont {Ma},
  \citenamefont {Eremets}, \citenamefont {Oganov}, \citenamefont {Xie},
  \citenamefont {Trojan}, \citenamefont {Medvedev}, \citenamefont {Lyakhov},
  \citenamefont {Valle},\ and\ \citenamefont {Prakapenka}}]{ma2009transparent}%
  \BibitemOpen
  \bibfield  {author} {\bibinfo {author} {\bibfnamefont {Y.}~\bibnamefont
  {Ma}}, \bibinfo {author} {\bibfnamefont {M.}~\bibnamefont {Eremets}},
  \bibinfo {author} {\bibfnamefont {A.~R.}\ \bibnamefont {Oganov}}, \bibinfo
  {author} {\bibfnamefont {Y.}~\bibnamefont {Xie}}, \bibinfo {author}
  {\bibfnamefont {I.}~\bibnamefont {Trojan}}, \bibinfo {author} {\bibfnamefont
  {S.}~\bibnamefont {Medvedev}}, \bibinfo {author} {\bibfnamefont {A.~O.}\
  \bibnamefont {Lyakhov}}, \bibinfo {author} {\bibfnamefont {M.}~\bibnamefont
  {Valle}}, \ and\ \bibinfo {author} {\bibfnamefont {V.}~\bibnamefont
  {Prakapenka}},\ }\href@noop {} {\bibfield  {journal} {\bibinfo  {journal}
  {Nature}\ }\textbf {\bibinfo {volume} {458}},\ \bibinfo {pages} {182}
  (\bibinfo {year} {2009})}\BibitemShut {NoStop}%
\bibitem [{\citenamefont {McMahon}\ and\ \citenamefont
  {Nelmes}(2006)}]{mcmahon2006high}%
  \BibitemOpen
  \bibfield  {author} {\bibinfo {author} {\bibfnamefont {M.~I.}\ \bibnamefont
  {McMahon}}\ and\ \bibinfo {author} {\bibfnamefont {R.~J.}\ \bibnamefont
  {Nelmes}},\ }\href@noop {} {\bibfield  {journal} {\bibinfo  {journal}
  {Chemical Society Reviews}\ }\textbf {\bibinfo {volume} {35}},\ \bibinfo
  {pages} {943} (\bibinfo {year} {2006})}\BibitemShut {NoStop}%
\bibitem [{\citenamefont {McMahon}\ and\ \citenamefont
  {Nelmes}(2004)}]{mcmahon2004incommensurate}%
  \BibitemOpen
  \bibfield  {author} {\bibinfo {author} {\bibfnamefont {M.}~\bibnamefont
  {McMahon}}\ and\ \bibinfo {author} {\bibfnamefont {R.}~\bibnamefont
  {Nelmes}},\ }\href@noop {} {\bibfield  {journal} {\bibinfo  {journal}
  {Zeitschrift f{\"u}r Kristallographie-Crystalline Materials}\ }\textbf
  {\bibinfo {volume} {219}},\ \bibinfo {pages} {742} (\bibinfo {year}
  {2004})}\BibitemShut {NoStop}%
\bibitem [{\citenamefont {Pickard}\ and\ \citenamefont
  {Needs}(2010)}]{pickard2010aluminium}%
  \BibitemOpen
  \bibfield  {author} {\bibinfo {author} {\bibfnamefont {C.~J.}\ \bibnamefont
  {Pickard}}\ and\ \bibinfo {author} {\bibfnamefont {R.}~\bibnamefont
  {Needs}},\ }\href@noop {} {\bibfield  {journal} {\bibinfo  {journal} {Nature
  Materials}\ }\textbf {\bibinfo {volume} {9}},\ \bibinfo {pages} {624}
  (\bibinfo {year} {2010})}\BibitemShut {NoStop}%
\bibitem [{\citenamefont {Gorman}\ \emph {et~al.}(2022)\citenamefont {Gorman},
  \citenamefont {Elatresh}, \citenamefont {Lazicki}, \citenamefont {Cormier},
  \citenamefont {Bonev}, \citenamefont {McGonegle}, \citenamefont {Briggs},
  \citenamefont {Coleman}, \citenamefont {Rothman}, \citenamefont {Peacock}
  \emph {et~al.}}]{gorman2022experimental}%
  \BibitemOpen
  \bibfield  {author} {\bibinfo {author} {\bibfnamefont {M.~G.}\ \bibnamefont
  {Gorman}}, \bibinfo {author} {\bibfnamefont {S.}~\bibnamefont {Elatresh}},
  \bibinfo {author} {\bibfnamefont {A.}~\bibnamefont {Lazicki}}, \bibinfo
  {author} {\bibfnamefont {M.~M.}\ \bibnamefont {Cormier}}, \bibinfo {author}
  {\bibfnamefont {S.}~\bibnamefont {Bonev}}, \bibinfo {author} {\bibfnamefont
  {D.}~\bibnamefont {McGonegle}}, \bibinfo {author} {\bibfnamefont
  {R.}~\bibnamefont {Briggs}}, \bibinfo {author} {\bibfnamefont
  {A.}~\bibnamefont {Coleman}}, \bibinfo {author} {\bibfnamefont
  {S.}~\bibnamefont {Rothman}}, \bibinfo {author} {\bibfnamefont
  {L.}~\bibnamefont {Peacock}},  \emph {et~al.},\ }\href@noop {} {\bibfield
  {journal} {\bibinfo  {journal} {Nature Physics}\ }\textbf {\bibinfo {volume}
  {18}},\ \bibinfo {pages} {1307} (\bibinfo {year} {2022})}\BibitemShut
  {NoStop}%
\bibitem [{\citenamefont {Pickard}\ and\ \citenamefont
  {Needs}(2011{\natexlab{b}})}]{pickard2011predicted}%
  \BibitemOpen
  \bibfield  {author} {\bibinfo {author} {\bibfnamefont {C.~J.}\ \bibnamefont
  {Pickard}}\ and\ \bibinfo {author} {\bibfnamefont {R.}~\bibnamefont
  {Needs}},\ }\href@noop {} {\bibfield  {journal} {\bibinfo  {journal}
  {Physical Review Letters}\ }\textbf {\bibinfo {volume} {107}},\ \bibinfo
  {pages} {087201} (\bibinfo {year} {2011}{\natexlab{b}})}\BibitemShut
  {NoStop}%
\bibitem [{\citenamefont {Garisto}(2024)}]{garisto2024superconductivity}%
  \BibitemOpen
  \bibfield  {author} {\bibinfo {author} {\bibfnamefont {D.}~\bibnamefont
  {Garisto}},\ }\href@noop {} {\bibfield  {journal} {\bibinfo  {journal}
  {Nature}\ } (\bibinfo {year} {2024})}\BibitemShut {NoStop}%
\bibitem [{\citenamefont {Clark}\ \emph {et~al.}(2005)\citenamefont {Clark},
  \citenamefont {Segall}, \citenamefont {Pickard}, \citenamefont {Hasnip},
  \citenamefont {Probert}, \citenamefont {Refson},\ and\ \citenamefont
  {Payne}}]{clark2005first}%
  \BibitemOpen
  \bibfield  {author} {\bibinfo {author} {\bibfnamefont {S.~J.}\ \bibnamefont
  {Clark}}, \bibinfo {author} {\bibfnamefont {M.~D.}\ \bibnamefont {Segall}},
  \bibinfo {author} {\bibfnamefont {C.~J.}\ \bibnamefont {Pickard}}, \bibinfo
  {author} {\bibfnamefont {P.~J.}\ \bibnamefont {Hasnip}}, \bibinfo {author}
  {\bibfnamefont {M.~I.}\ \bibnamefont {Probert}}, \bibinfo {author}
  {\bibfnamefont {K.}~\bibnamefont {Refson}}, \ and\ \bibinfo {author}
  {\bibfnamefont {M.~C.}\ \bibnamefont {Payne}},\ }\href@noop {} {\bibfield
  {journal} {\bibinfo  {journal} {Zeitschrift f{\"u}r
  Kristallographie-Crystalline Materials}\ }\textbf {\bibinfo {volume} {220}},\
  \bibinfo {pages} {567} (\bibinfo {year} {2005})}\BibitemShut {NoStop}%
\bibitem [{\citenamefont {Conway}\ \emph {et~al.}(2021)\citenamefont {Conway},
  \citenamefont {Pickard},\ and\ \citenamefont {Hermann}}]{conway2021rules}%
  \BibitemOpen
  \bibfield  {author} {\bibinfo {author} {\bibfnamefont {L.~J.}\ \bibnamefont
  {Conway}}, \bibinfo {author} {\bibfnamefont {C.~J.}\ \bibnamefont {Pickard}},
  \ and\ \bibinfo {author} {\bibfnamefont {A.}~\bibnamefont {Hermann}},\
  }\href@noop {} {\bibfield  {journal} {\bibinfo  {journal} {Proceedings of the
  National Academy of Sciences}\ }\textbf {\bibinfo {volume} {118}},\ \bibinfo
  {pages} {e2026360118} (\bibinfo {year} {2021})}\BibitemShut {NoStop}%
\bibitem [{\citenamefont {Nelson}\ \emph {et~al.}(2021)\citenamefont {Nelson},
  \citenamefont {Needs},\ and\ \citenamefont {Pickard}}]{nelson2021navigating}%
  \BibitemOpen
  \bibfield  {author} {\bibinfo {author} {\bibfnamefont {J.~R.}\ \bibnamefont
  {Nelson}}, \bibinfo {author} {\bibfnamefont {R.~J.}\ \bibnamefont {Needs}}, \
  and\ \bibinfo {author} {\bibfnamefont {C.~J.}\ \bibnamefont {Pickard}},\
  }\href@noop {} {\bibfield  {journal} {\bibinfo  {journal} {Physical Review
  Materials}\ }\textbf {\bibinfo {volume} {5}},\ \bibinfo {pages} {123801}
  (\bibinfo {year} {2021})}\BibitemShut {NoStop}%
\bibitem [{\citenamefont {Shipley}\ \emph {et~al.}(2021)\citenamefont
  {Shipley}, \citenamefont {Hutcheon}, \citenamefont {Needs},\ and\
  \citenamefont {Pickard}}]{shipley2021high}%
  \BibitemOpen
  \bibfield  {author} {\bibinfo {author} {\bibfnamefont {A.~M.}\ \bibnamefont
  {Shipley}}, \bibinfo {author} {\bibfnamefont {M.~J.}\ \bibnamefont
  {Hutcheon}}, \bibinfo {author} {\bibfnamefont {R.~J.}\ \bibnamefont {Needs}},
  \ and\ \bibinfo {author} {\bibfnamefont {C.~J.}\ \bibnamefont {Pickard}},\
  }\href@noop {} {\bibfield  {journal} {\bibinfo  {journal} {Physical Review
  B}\ }\textbf {\bibinfo {volume} {104}},\ \bibinfo {pages} {054501} (\bibinfo
  {year} {2021})}\BibitemShut {NoStop}%
\bibitem [{\citenamefont {Dolui}\ \emph {et~al.}(2024)\citenamefont {Dolui},
  \citenamefont {Conway}, \citenamefont {Heil}, \citenamefont {Strobel},
  \citenamefont {Prasankumar},\ and\ \citenamefont
  {Pickard}}]{dolui2024feasible}%
  \BibitemOpen
  \bibfield  {author} {\bibinfo {author} {\bibfnamefont {K.}~\bibnamefont
  {Dolui}}, \bibinfo {author} {\bibfnamefont {L.~J.}\ \bibnamefont {Conway}},
  \bibinfo {author} {\bibfnamefont {C.}~\bibnamefont {Heil}}, \bibinfo {author}
  {\bibfnamefont {T.~A.}\ \bibnamefont {Strobel}}, \bibinfo {author}
  {\bibfnamefont {R.~P.}\ \bibnamefont {Prasankumar}}, \ and\ \bibinfo {author}
  {\bibfnamefont {C.~J.}\ \bibnamefont {Pickard}},\ }\href@noop {} {\bibfield
  {journal} {\bibinfo  {journal} {Physical Review Letters}\ }\textbf {\bibinfo
  {volume} {132}},\ \bibinfo {pages} {166001} (\bibinfo {year}
  {2024})}\BibitemShut {NoStop}%
\bibitem [{AIR()}]{AIRSS-website}%
  \BibitemOpen
  \href@noop {} {}\bibinfo {howpublished}
  {\url{https://www.mtg.msm.cam.ac.uk/Codes/AIRSS}}\BibitemShut {NoStop}%
\bibitem [{\citenamefont {Prechelt}(1998)}]{prechelt1998early}%
  \BibitemOpen
  \bibfield  {author} {\bibinfo {author} {\bibfnamefont {L.}~\bibnamefont
  {Prechelt}},\ }in\ \href@noop {} {\emph {\bibinfo {booktitle} {Neural
  Networks: Tricks of the trade}}}\ (\bibinfo  {publisher} {Springer},\
  \bibinfo {year} {1998})\ pp.\ \bibinfo {pages} {55--69}\BibitemShut {NoStop}%
\bibitem [{\citenamefont {Hansen}\ and\ \citenamefont
  {Salamon}(1990)}]{hansen1990neural}%
  \BibitemOpen
  \bibfield  {author} {\bibinfo {author} {\bibfnamefont {L.~K.}\ \bibnamefont
  {Hansen}}\ and\ \bibinfo {author} {\bibfnamefont {P.}~\bibnamefont
  {Salamon}},\ }\href@noop {} {\bibfield  {journal} {\bibinfo  {journal} {IEEE
  Transactions on Pattern Analysis and Machine Intelligence}\ }\textbf
  {\bibinfo {volume} {12}},\ \bibinfo {pages} {993} (\bibinfo {year}
  {1990})}\BibitemShut {NoStop}%
\bibitem [{\citenamefont {Schran}\ \emph {et~al.}(2020)\citenamefont {Schran},
  \citenamefont {Brezina},\ and\ \citenamefont
  {Marsalek}}]{schran2020committee}%
  \BibitemOpen
  \bibfield  {author} {\bibinfo {author} {\bibfnamefont {C.}~\bibnamefont
  {Schran}}, \bibinfo {author} {\bibfnamefont {K.}~\bibnamefont {Brezina}}, \
  and\ \bibinfo {author} {\bibfnamefont {O.}~\bibnamefont {Marsalek}},\
  }\href@noop {} {\bibfield  {journal} {\bibinfo  {journal} {The Journal of
  Chemical Physics}\ }\textbf {\bibinfo {volume} {153}},\ \bibinfo {pages}
  {104105} (\bibinfo {year} {2020})}\BibitemShut {NoStop}%
\bibitem [{\citenamefont {Lopanitsyna}\ \emph {et~al.}(2023)\citenamefont
  {Lopanitsyna}, \citenamefont {Fraux}, \citenamefont {Springer}, \citenamefont
  {De},\ and\ \citenamefont {Ceriotti}}]{lopanitsyna2023modeling}%
  \BibitemOpen
  \bibfield  {author} {\bibinfo {author} {\bibfnamefont {N.}~\bibnamefont
  {Lopanitsyna}}, \bibinfo {author} {\bibfnamefont {G.}~\bibnamefont {Fraux}},
  \bibinfo {author} {\bibfnamefont {M.~A.}\ \bibnamefont {Springer}}, \bibinfo
  {author} {\bibfnamefont {S.}~\bibnamefont {De}}, \ and\ \bibinfo {author}
  {\bibfnamefont {M.}~\bibnamefont {Ceriotti}},\ }\href@noop {} {\bibfield
  {journal} {\bibinfo  {journal} {Physical Review Materials}\ }\textbf
  {\bibinfo {volume} {7}},\ \bibinfo {pages} {045802} (\bibinfo {year}
  {2023})}\BibitemShut {NoStop}%
\bibitem [{EDD()}]{EDDP-website}%
  \BibitemOpen
  \href@noop {} {}\bibinfo {howpublished}
  {\url{https://www.mtg.msm.cam.ac.uk/Codes/EDDP}}\BibitemShut {NoStop}%
\bibitem [{\citenamefont {Oganov}\ \emph {et~al.}(2009)\citenamefont {Oganov},
  \citenamefont {Chen}, \citenamefont {Gatti}, \citenamefont {Ma},
  \citenamefont {Ma}, \citenamefont {Glass}, \citenamefont {Liu}, \citenamefont
  {Yu}, \citenamefont {Kurakevych},\ and\ \citenamefont
  {Solozhenko}}]{oganov2009ionic}%
  \BibitemOpen
  \bibfield  {author} {\bibinfo {author} {\bibfnamefont {A.~R.}\ \bibnamefont
  {Oganov}}, \bibinfo {author} {\bibfnamefont {J.}~\bibnamefont {Chen}},
  \bibinfo {author} {\bibfnamefont {C.}~\bibnamefont {Gatti}}, \bibinfo
  {author} {\bibfnamefont {Y.}~\bibnamefont {Ma}}, \bibinfo {author}
  {\bibfnamefont {Y.}~\bibnamefont {Ma}}, \bibinfo {author} {\bibfnamefont
  {C.~W.}\ \bibnamefont {Glass}}, \bibinfo {author} {\bibfnamefont
  {Z.}~\bibnamefont {Liu}}, \bibinfo {author} {\bibfnamefont {T.}~\bibnamefont
  {Yu}}, \bibinfo {author} {\bibfnamefont {O.~O.}\ \bibnamefont {Kurakevych}},
  \ and\ \bibinfo {author} {\bibfnamefont {V.~L.}\ \bibnamefont {Solozhenko}},\
  }\href@noop {} {\bibfield  {journal} {\bibinfo  {journal} {Nature}\ }\textbf
  {\bibinfo {volume} {457}},\ \bibinfo {pages} {863} (\bibinfo {year}
  {2009})}\BibitemShut {NoStop}%
\bibitem [{\citenamefont {Cheng}\ \emph {et~al.}(2020)\citenamefont {Cheng},
  \citenamefont {Mazzola}, \citenamefont {Pickard},\ and\ \citenamefont
  {Ceriotti}}]{cheng2020evidence}%
  \BibitemOpen
  \bibfield  {author} {\bibinfo {author} {\bibfnamefont {B.}~\bibnamefont
  {Cheng}}, \bibinfo {author} {\bibfnamefont {G.}~\bibnamefont {Mazzola}},
  \bibinfo {author} {\bibfnamefont {C.~J.}\ \bibnamefont {Pickard}}, \ and\
  \bibinfo {author} {\bibfnamefont {M.}~\bibnamefont {Ceriotti}},\ }\href@noop
  {} {\bibfield  {journal} {\bibinfo  {journal} {Nature}\ }\textbf {\bibinfo
  {volume} {585}},\ \bibinfo {pages} {217} (\bibinfo {year}
  {2020})}\BibitemShut {NoStop}%
\bibitem [{\citenamefont {Schran}\ \emph {et~al.}(2021)\citenamefont {Schran},
  \citenamefont {Thiemann}, \citenamefont {Rowe}, \citenamefont {M{\"u}ller},
  \citenamefont {Marsalek},\ and\ \citenamefont
  {Michaelides}}]{schran2021machine}%
  \BibitemOpen
  \bibfield  {author} {\bibinfo {author} {\bibfnamefont {C.}~\bibnamefont
  {Schran}}, \bibinfo {author} {\bibfnamefont {F.~L.}\ \bibnamefont
  {Thiemann}}, \bibinfo {author} {\bibfnamefont {P.}~\bibnamefont {Rowe}},
  \bibinfo {author} {\bibfnamefont {E.~A.}\ \bibnamefont {M{\"u}ller}},
  \bibinfo {author} {\bibfnamefont {O.}~\bibnamefont {Marsalek}}, \ and\
  \bibinfo {author} {\bibfnamefont {A.}~\bibnamefont {Michaelides}},\
  }\href@noop {} {\bibfield  {journal} {\bibinfo  {journal} {Proceedings of the
  National Academy of Sciences}\ }\textbf {\bibinfo {volume} {118}},\ \bibinfo
  {pages} {e2110077118} (\bibinfo {year} {2021})}\BibitemShut {NoStop}%
\bibitem [{\citenamefont {Deringer}\ \emph {et~al.}(2021)\citenamefont
  {Deringer}, \citenamefont {Bernstein}, \citenamefont {Cs{\'a}nyi},
  \citenamefont {Ben~Mahmoud}, \citenamefont {Ceriotti}, \citenamefont
  {Wilson}, \citenamefont {Drabold},\ and\ \citenamefont
  {Elliott}}]{deringer2021origins}%
  \BibitemOpen
  \bibfield  {author} {\bibinfo {author} {\bibfnamefont {V.~L.}\ \bibnamefont
  {Deringer}}, \bibinfo {author} {\bibfnamefont {N.}~\bibnamefont {Bernstein}},
  \bibinfo {author} {\bibfnamefont {G.}~\bibnamefont {Cs{\'a}nyi}}, \bibinfo
  {author} {\bibfnamefont {C.}~\bibnamefont {Ben~Mahmoud}}, \bibinfo {author}
  {\bibfnamefont {M.}~\bibnamefont {Ceriotti}}, \bibinfo {author}
  {\bibfnamefont {M.}~\bibnamefont {Wilson}}, \bibinfo {author} {\bibfnamefont
  {D.~A.}\ \bibnamefont {Drabold}}, \ and\ \bibinfo {author} {\bibfnamefont
  {S.~R.}\ \bibnamefont {Elliott}},\ }\href@noop {} {\bibfield  {journal}
  {\bibinfo  {journal} {Nature}\ }\textbf {\bibinfo {volume} {589}},\ \bibinfo
  {pages} {59} (\bibinfo {year} {2021})}\BibitemShut {NoStop}%
\bibitem [{\citenamefont {Perdew}\ \emph {et~al.}(1996)\citenamefont {Perdew},
  \citenamefont {Burke},\ and\ \citenamefont
  {Ernzerhof}}]{perdew1996generalized}%
  \BibitemOpen
  \bibfield  {author} {\bibinfo {author} {\bibfnamefont {J.~P.}\ \bibnamefont
  {Perdew}}, \bibinfo {author} {\bibfnamefont {K.}~\bibnamefont {Burke}}, \
  and\ \bibinfo {author} {\bibfnamefont {M.}~\bibnamefont {Ernzerhof}},\
  }\href@noop {} {\bibfield  {journal} {\bibinfo  {journal} {Physical Review
  Letters}\ }\textbf {\bibinfo {volume} {77}},\ \bibinfo {pages} {3865}
  (\bibinfo {year} {1996})}\BibitemShut {NoStop}%
\bibitem [{\citenamefont {Ahnert}\ \emph {et~al.}(2017)\citenamefont {Ahnert},
  \citenamefont {Grant},\ and\ \citenamefont {Pickard}}]{ahnert2017revealing}%
  \BibitemOpen
  \bibfield  {author} {\bibinfo {author} {\bibfnamefont {S.~E.}\ \bibnamefont
  {Ahnert}}, \bibinfo {author} {\bibfnamefont {W.~P.}\ \bibnamefont {Grant}}, \
  and\ \bibinfo {author} {\bibfnamefont {C.~J.}\ \bibnamefont {Pickard}},\
  }\href@noop {} {\bibfield  {journal} {\bibinfo  {journal} {NPJ Computational
  Materials}\ }\textbf {\bibinfo {volume} {3}},\ \bibinfo {pages} {1} (\bibinfo
  {year} {2017})}\BibitemShut {NoStop}%
\bibitem [{\citenamefont {Callmer}(1977)}]{callmer1977accurate}%
  \BibitemOpen
  \bibfield  {author} {\bibinfo {author} {\bibfnamefont {B.}~\bibnamefont
  {Callmer}},\ }\href@noop {} {\bibfield  {journal} {\bibinfo  {journal} {Acta
  Crystallographica Section B: Structural Crystallography and Crystal
  Chemistry}\ }\textbf {\bibinfo {volume} {33}},\ \bibinfo {pages} {1951}
  (\bibinfo {year} {1977})}\BibitemShut {NoStop}%
\bibitem [{\citenamefont {Van~Setten}\ \emph {et~al.}(2007)\citenamefont
  {Van~Setten}, \citenamefont {Uijttewaal}, \citenamefont {de~Wijs},\ and\
  \citenamefont {de~Groot}}]{van2007thermodynamic}%
  \BibitemOpen
  \bibfield  {author} {\bibinfo {author} {\bibfnamefont {M.~J.}\ \bibnamefont
  {Van~Setten}}, \bibinfo {author} {\bibfnamefont {M.~A.}\ \bibnamefont
  {Uijttewaal}}, \bibinfo {author} {\bibfnamefont {G.~A.}\ \bibnamefont
  {de~Wijs}}, \ and\ \bibinfo {author} {\bibfnamefont {R.~A.}\ \bibnamefont
  {de~Groot}},\ }\href@noop {} {\bibfield  {journal} {\bibinfo  {journal}
  {Journal of the American Chemical Society}\ }\textbf {\bibinfo {volume}
  {129}},\ \bibinfo {pages} {2458} (\bibinfo {year} {2007})}\BibitemShut
  {NoStop}%
\bibitem [{\citenamefont {Podryabinkin}\ \emph {et~al.}(2019)\citenamefont
  {Podryabinkin}, \citenamefont {Tikhonov}, \citenamefont {Shapeev},\ and\
  \citenamefont {Oganov}}]{podryabinkin2019accelerating}%
  \BibitemOpen
  \bibfield  {author} {\bibinfo {author} {\bibfnamefont {E.~V.}\ \bibnamefont
  {Podryabinkin}}, \bibinfo {author} {\bibfnamefont {E.~V.}\ \bibnamefont
  {Tikhonov}}, \bibinfo {author} {\bibfnamefont {A.~V.}\ \bibnamefont
  {Shapeev}}, \ and\ \bibinfo {author} {\bibfnamefont {A.~R.}\ \bibnamefont
  {Oganov}},\ }\href@noop {} {\bibfield  {journal} {\bibinfo  {journal}
  {Physical Review B}\ }\textbf {\bibinfo {volume} {99}},\ \bibinfo {pages}
  {064114} (\bibinfo {year} {2019})}\BibitemShut {NoStop}%
\bibitem [{\citenamefont {Shapeev}(2016)}]{shapeev2016moment}%
  \BibitemOpen
  \bibfield  {author} {\bibinfo {author} {\bibfnamefont {A.~V.}\ \bibnamefont
  {Shapeev}},\ }\href@noop {} {\bibfield  {journal} {\bibinfo  {journal}
  {Multiscale Modeling \& Simulation}\ }\textbf {\bibinfo {volume} {14}},\
  \bibinfo {pages} {1153} (\bibinfo {year} {2016})}\BibitemShut {NoStop}%
\bibitem [{\citenamefont {Hayami}\ \emph {et~al.}(2024)\citenamefont {Hayami},
  \citenamefont {Hiroto}, \citenamefont {Soga}, \citenamefont {Ogitsu},\ and\
  \citenamefont {Kimura}}]{hayami2024thermodynamic}%
  \BibitemOpen
  \bibfield  {author} {\bibinfo {author} {\bibfnamefont {W.}~\bibnamefont
  {Hayami}}, \bibinfo {author} {\bibfnamefont {T.}~\bibnamefont {Hiroto}},
  \bibinfo {author} {\bibfnamefont {K.}~\bibnamefont {Soga}}, \bibinfo {author}
  {\bibfnamefont {T.}~\bibnamefont {Ogitsu}}, \ and\ \bibinfo {author}
  {\bibfnamefont {K.}~\bibnamefont {Kimura}},\ }\href@noop {} {\bibfield
  {journal} {\bibinfo  {journal} {Journal of Solid State Chemistry}\ }\textbf
  {\bibinfo {volume} {329}},\ \bibinfo {pages} {124407} (\bibinfo {year}
  {2024})}\BibitemShut {NoStop}%
\bibitem [{\citenamefont {Hayami}(2015)}]{hayami2015structural}%
  \BibitemOpen
  \bibfield  {author} {\bibinfo {author} {\bibfnamefont {W.}~\bibnamefont
  {Hayami}},\ }\href@noop {} {\bibfield  {journal} {\bibinfo  {journal}
  {Journal of Solid State Chemistry}\ }\textbf {\bibinfo {volume} {221}},\
  \bibinfo {pages} {378} (\bibinfo {year} {2015})}\BibitemShut {NoStop}%
\bibitem [{\citenamefont {Doll}\ \emph {et~al.}(2008)\citenamefont {Doll},
  \citenamefont {Sch{\"o}n},\ and\ \citenamefont {Jansen}}]{doll2008structure}%
  \BibitemOpen
  \bibfield  {author} {\bibinfo {author} {\bibfnamefont {K.}~\bibnamefont
  {Doll}}, \bibinfo {author} {\bibfnamefont {J.}~\bibnamefont {Sch{\"o}n}}, \
  and\ \bibinfo {author} {\bibfnamefont {M.}~\bibnamefont {Jansen}},\ }in\
  \href@noop {} {\emph {\bibinfo {booktitle} {Journal of Physics: Conference
  Series}}},\ Vol.\ \bibinfo {volume} {117}\ (\bibinfo {organization} {IOP
  Publishing},\ \bibinfo {year} {2008})\ p.\ \bibinfo {pages}
  {012014}\BibitemShut {NoStop}%
\bibitem [{\citenamefont {Wales}\ and\ \citenamefont
  {Doye}(1997)}]{wales1997global}%
  \BibitemOpen
  \bibfield  {author} {\bibinfo {author} {\bibfnamefont {D.~J.}\ \bibnamefont
  {Wales}}\ and\ \bibinfo {author} {\bibfnamefont {J.~P.}\ \bibnamefont
  {Doye}},\ }\href@noop {} {\bibfield  {journal} {\bibinfo  {journal} {The
  Journal of Physical Chemistry A}\ }\textbf {\bibinfo {volume} {101}},\
  \bibinfo {pages} {5111} (\bibinfo {year} {1997})}\BibitemShut {NoStop}%
\bibitem [{\citenamefont {Goedecker}(2004)}]{goedecker2004minima}%
  \BibitemOpen
  \bibfield  {author} {\bibinfo {author} {\bibfnamefont {S.}~\bibnamefont
  {Goedecker}},\ }\href@noop {} {\bibfield  {journal} {\bibinfo  {journal} {The
  Journal of chemical physics}\ }\textbf {\bibinfo {volume} {120}},\ \bibinfo
  {pages} {9911} (\bibinfo {year} {2004})}\BibitemShut {NoStop}%
\bibitem [{\citenamefont {Hoffmann}\ \emph {et~al.}(2016)\citenamefont
  {Hoffmann}, \citenamefont {Kabanov}, \citenamefont {Golov},\ and\
  \citenamefont {Proserpio}}]{hoffmann2016homo}%
  \BibitemOpen
  \bibfield  {author} {\bibinfo {author} {\bibfnamefont {R.}~\bibnamefont
  {Hoffmann}}, \bibinfo {author} {\bibfnamefont {A.~A.}\ \bibnamefont
  {Kabanov}}, \bibinfo {author} {\bibfnamefont {A.~A.}\ \bibnamefont {Golov}},
  \ and\ \bibinfo {author} {\bibfnamefont {D.~M.}\ \bibnamefont {Proserpio}},\
  }\href@noop {} {\bibfield  {journal} {\bibinfo  {journal} {Angewandte Chemie
  International Edition}\ }\textbf {\bibinfo {volume} {55}},\ \bibinfo {pages}
  {10962} (\bibinfo {year} {2016})}\BibitemShut {NoStop}%
\bibitem [{\citenamefont {Yin}(1984)}]{yin1984si}%
  \BibitemOpen
  \bibfield  {author} {\bibinfo {author} {\bibfnamefont {M.}~\bibnamefont
  {Yin}},\ }\href@noop {} {\bibfield  {journal} {\bibinfo  {journal} {Physical
  Review B}\ }\textbf {\bibinfo {volume} {30}},\ \bibinfo {pages} {1773}
  (\bibinfo {year} {1984})}\BibitemShut {NoStop}%
\bibitem [{\citenamefont {Grumbach}\ and\ \citenamefont
  {Martin}(1996)}]{grumbach1996phase}%
  \BibitemOpen
  \bibfield  {author} {\bibinfo {author} {\bibfnamefont {M.~P.}\ \bibnamefont
  {Grumbach}}\ and\ \bibinfo {author} {\bibfnamefont {R.~M.}\ \bibnamefont
  {Martin}},\ }\href@noop {} {\bibfield  {journal} {\bibinfo  {journal}
  {Physical review B}\ }\textbf {\bibinfo {volume} {54}},\ \bibinfo {pages}
  {15730} (\bibinfo {year} {1996})}\BibitemShut {NoStop}%
\bibitem [{\citenamefont {Martinez-Canales}\ \emph {et~al.}(2012)\citenamefont
  {Martinez-Canales}, \citenamefont {Pickard},\ and\ \citenamefont
  {Needs}}]{martinez2012thermodynamically}%
  \BibitemOpen
  \bibfield  {author} {\bibinfo {author} {\bibfnamefont {M.}~\bibnamefont
  {Martinez-Canales}}, \bibinfo {author} {\bibfnamefont {C.~J.}\ \bibnamefont
  {Pickard}}, \ and\ \bibinfo {author} {\bibfnamefont {R.~J.}\ \bibnamefont
  {Needs}},\ }\href@noop {} {\bibfield  {journal} {\bibinfo  {journal}
  {Physical Review Letters}\ }\textbf {\bibinfo {volume} {108}},\ \bibinfo
  {pages} {045704} (\bibinfo {year} {2012})}\BibitemShut {NoStop}%
\bibitem [{\citenamefont {Lu}(2022)}]{Lu2022}%
  \BibitemOpen
  \bibfield  {author} {\bibinfo {author} {\bibfnamefont {X.}~\bibnamefont
  {Lu}},\ }\enquote {\bibinfo {title} {Connecting fullerenes with carbon
  nanotubes and graphene},}\ in\ \href {\doibase 10.1007/978-981-16-8994-9_8}
  {\emph {\bibinfo {booktitle} {Handbook of Fullerene Science and
  Technology}}},\ \bibinfo {editor} {edited by\ \bibinfo {editor}
  {\bibfnamefont {X.}~\bibnamefont {Lu}}, \bibinfo {editor} {\bibfnamefont
  {T.}~\bibnamefont {Akasaka}}, \ and\ \bibinfo {editor} {\bibfnamefont
  {Z.}~\bibnamefont {Slanina}}}\ (\bibinfo  {publisher} {Springer Nature
  Singapore},\ \bibinfo {address} {Singapore},\ \bibinfo {year} {2022})\ pp.\
  \bibinfo {pages} {265--270}\BibitemShut {NoStop}%
\bibitem [{\citenamefont {N{\'e}meth}\ \emph {et~al.}(2021)\citenamefont
  {N{\'e}meth}, \citenamefont {McColl}, \citenamefont {Garvie}, \citenamefont
  {Salzmann}, \citenamefont {Pickard}, \citenamefont {Cora}, \citenamefont
  {Smith}, \citenamefont {Mezouar}, \citenamefont {Howard},\ and\ \citenamefont
  {McMillan}}]{nemeth2021diaphite}%
  \BibitemOpen
  \bibfield  {author} {\bibinfo {author} {\bibfnamefont {P.}~\bibnamefont
  {N{\'e}meth}}, \bibinfo {author} {\bibfnamefont {K.}~\bibnamefont {McColl}},
  \bibinfo {author} {\bibfnamefont {L.~A.}\ \bibnamefont {Garvie}}, \bibinfo
  {author} {\bibfnamefont {C.~G.}\ \bibnamefont {Salzmann}}, \bibinfo {author}
  {\bibfnamefont {C.~J.}\ \bibnamefont {Pickard}}, \bibinfo {author}
  {\bibfnamefont {F.}~\bibnamefont {Cora}}, \bibinfo {author} {\bibfnamefont
  {R.~L.}\ \bibnamefont {Smith}}, \bibinfo {author} {\bibfnamefont
  {M.}~\bibnamefont {Mezouar}}, \bibinfo {author} {\bibfnamefont {C.~A.}\
  \bibnamefont {Howard}}, \ and\ \bibinfo {author} {\bibfnamefont {P.~F.}\
  \bibnamefont {McMillan}},\ }\href@noop {} {\bibfield  {journal} {\bibinfo
  {journal} {Diamond and Related Materials}\ }\textbf {\bibinfo {volume}
  {119}},\ \bibinfo {pages} {108573} (\bibinfo {year} {2021})}\BibitemShut
  {NoStop}%
\bibitem [{\citenamefont {Kim}\ \emph {et~al.}(2020)\citenamefont {Kim},
  \citenamefont {Noh}, \citenamefont {Gu}, \citenamefont {Aspuru-Guzik},\ and\
  \citenamefont {Jung}}]{kim2020generative}%
  \BibitemOpen
  \bibfield  {author} {\bibinfo {author} {\bibfnamefont {S.}~\bibnamefont
  {Kim}}, \bibinfo {author} {\bibfnamefont {J.}~\bibnamefont {Noh}}, \bibinfo
  {author} {\bibfnamefont {G.~H.}\ \bibnamefont {Gu}}, \bibinfo {author}
  {\bibfnamefont {A.}~\bibnamefont {Aspuru-Guzik}}, \ and\ \bibinfo {author}
  {\bibfnamefont {Y.}~\bibnamefont {Jung}},\ }\href@noop {} {\bibfield
  {journal} {\bibinfo  {journal} {ACS central science}\ }\textbf {\bibinfo
  {volume} {6}},\ \bibinfo {pages} {1412} (\bibinfo {year} {2020})}\BibitemShut
  {NoStop}%
\bibitem [{\citenamefont {Court}\ \emph {et~al.}(2020)\citenamefont {Court},
  \citenamefont {Yildirim}, \citenamefont {Jain},\ and\ \citenamefont
  {Cole}}]{court20203}%
  \BibitemOpen
  \bibfield  {author} {\bibinfo {author} {\bibfnamefont {C.~J.}\ \bibnamefont
  {Court}}, \bibinfo {author} {\bibfnamefont {B.}~\bibnamefont {Yildirim}},
  \bibinfo {author} {\bibfnamefont {A.}~\bibnamefont {Jain}}, \ and\ \bibinfo
  {author} {\bibfnamefont {J.~M.}\ \bibnamefont {Cole}},\ }\href@noop {}
  {\bibfield  {journal} {\bibinfo  {journal} {Journal of Chemical Information
  and Modeling}\ }\textbf {\bibinfo {volume} {60}},\ \bibinfo {pages} {4518}
  (\bibinfo {year} {2020})}\BibitemShut {NoStop}%
\bibitem [{\citenamefont {Pakornchote}\ \emph {et~al.}(2024)\citenamefont
  {Pakornchote}, \citenamefont {Choomphon-Anomakhun}, \citenamefont {Arrerut},
  \citenamefont {Atthapak}, \citenamefont {Khamkaeo}, \citenamefont
  {Chotibut},\ and\ \citenamefont
  {Bovornratanaraks}}]{pakornchote2024diffusion}%
  \BibitemOpen
  \bibfield  {author} {\bibinfo {author} {\bibfnamefont {T.}~\bibnamefont
  {Pakornchote}}, \bibinfo {author} {\bibfnamefont {N.}~\bibnamefont
  {Choomphon-Anomakhun}}, \bibinfo {author} {\bibfnamefont {S.}~\bibnamefont
  {Arrerut}}, \bibinfo {author} {\bibfnamefont {C.}~\bibnamefont {Atthapak}},
  \bibinfo {author} {\bibfnamefont {S.}~\bibnamefont {Khamkaeo}}, \bibinfo
  {author} {\bibfnamefont {T.}~\bibnamefont {Chotibut}}, \ and\ \bibinfo
  {author} {\bibfnamefont {T.}~\bibnamefont {Bovornratanaraks}},\ }\href@noop
  {} {\bibfield  {journal} {\bibinfo  {journal} {Scientific Reports}\ }\textbf
  {\bibinfo {volume} {14}},\ \bibinfo {pages} {1275} (\bibinfo {year}
  {2024})}\BibitemShut {NoStop}%
\bibitem [{\citenamefont {Luo}\ \emph {et~al.}(2024)\citenamefont {Luo},
  \citenamefont {Wang}, \citenamefont {Gao}, \citenamefont {Lv}, \citenamefont
  {Wang}, \citenamefont {Chen},\ and\ \citenamefont {Ma}}]{luo2024deep}%
  \BibitemOpen
  \bibfield  {author} {\bibinfo {author} {\bibfnamefont {X.}~\bibnamefont
  {Luo}}, \bibinfo {author} {\bibfnamefont {Z.}~\bibnamefont {Wang}}, \bibinfo
  {author} {\bibfnamefont {P.}~\bibnamefont {Gao}}, \bibinfo {author}
  {\bibfnamefont {J.}~\bibnamefont {Lv}}, \bibinfo {author} {\bibfnamefont
  {Y.}~\bibnamefont {Wang}}, \bibinfo {author} {\bibfnamefont {C.}~\bibnamefont
  {Chen}}, \ and\ \bibinfo {author} {\bibfnamefont {Y.}~\bibnamefont {Ma}},\
  }\href@noop {} {\bibfield  {journal} {\bibinfo  {journal} {arXiv preprint
  arXiv:2403.10846}\ } (\bibinfo {year} {2024})}\BibitemShut {NoStop}%
\bibitem [{\citenamefont {Cheng}(2024)}]{cheng2024response}%
  \BibitemOpen
  \bibfield  {author} {\bibinfo {author} {\bibfnamefont {B.}~\bibnamefont
  {Cheng}},\ }\href@noop {} {\bibfield  {journal} {\bibinfo  {journal} {arXiv
  preprint arXiv:2405.09057}\ } (\bibinfo {year} {2024})}\BibitemShut {NoStop}%
\bibitem [{\citenamefont {R{\o}nne}\ \emph {et~al.}(2024)\citenamefont
  {R{\o}nne}, \citenamefont {Aspuru-Guzik},\ and\ \citenamefont
  {Hammer}}]{ronne2024generative}%
  \BibitemOpen
  \bibfield  {author} {\bibinfo {author} {\bibfnamefont {N.}~\bibnamefont
  {R{\o}nne}}, \bibinfo {author} {\bibfnamefont {A.}~\bibnamefont
  {Aspuru-Guzik}}, \ and\ \bibinfo {author} {\bibfnamefont {B.}~\bibnamefont
  {Hammer}},\ }\href@noop {} {\bibfield  {journal} {\bibinfo  {journal} {arXiv
  preprint arXiv:2402.17404}\ } (\bibinfo {year} {2024})}\BibitemShut {NoStop}%
\bibitem [{\citenamefont {Permenter}\ and\ \citenamefont
  {Yuan}(2023)}]{permenter2023interpreting}%
  \BibitemOpen
  \bibfield  {author} {\bibinfo {author} {\bibfnamefont {F.}~\bibnamefont
  {Permenter}}\ and\ \bibinfo {author} {\bibfnamefont {C.}~\bibnamefont
  {Yuan}},\ }\href@noop {} {\bibfield  {journal} {\bibinfo  {journal} {arXiv
  preprint arXiv:2306.04848}\ } (\bibinfo {year} {2023})}\BibitemShut {NoStop}%
\bibitem [{\citenamefont {Batatia}\ \emph {et~al.}(2023)\citenamefont
  {Batatia}, \citenamefont {Benner}, \citenamefont {Chiang}, \citenamefont
  {Elena}, \citenamefont {Kov{\'a}cs}, \citenamefont {Riebesell}, \citenamefont
  {Advincula}, \citenamefont {Asta}, \citenamefont {Baldwin}, \citenamefont
  {Bernstein} \emph {et~al.}}]{batatia2023foundation}%
  \BibitemOpen
  \bibfield  {author} {\bibinfo {author} {\bibfnamefont {I.}~\bibnamefont
  {Batatia}}, \bibinfo {author} {\bibfnamefont {P.}~\bibnamefont {Benner}},
  \bibinfo {author} {\bibfnamefont {Y.}~\bibnamefont {Chiang}}, \bibinfo
  {author} {\bibfnamefont {A.~M.}\ \bibnamefont {Elena}}, \bibinfo {author}
  {\bibfnamefont {D.~P.}\ \bibnamefont {Kov{\'a}cs}}, \bibinfo {author}
  {\bibfnamefont {J.}~\bibnamefont {Riebesell}}, \bibinfo {author}
  {\bibfnamefont {X.~R.}\ \bibnamefont {Advincula}}, \bibinfo {author}
  {\bibfnamefont {M.}~\bibnamefont {Asta}}, \bibinfo {author} {\bibfnamefont
  {W.~J.}\ \bibnamefont {Baldwin}}, \bibinfo {author} {\bibfnamefont
  {N.}~\bibnamefont {Bernstein}},  \emph {et~al.},\ }\href@noop {} {\bibfield
  {journal} {\bibinfo  {journal} {arXiv preprint arXiv:2401.00096}\ } (\bibinfo
  {year} {2023})}\BibitemShut {NoStop}%
\end{thebibliography}%

\end{document}